\documentclass[a4paper,11pt]{article}
\pdfoutput=1 
\usepackage{jheppub} 
\usepackage{hyperref}
\usepackage{booktabs}
\usepackage[T1]{fontenc} 
\usepackage{siunitx} 
\usepackage{graphicx}%
\usepackage[dvipsnames]{xcolor}
\usepackage{dcolumn}
\usepackage{bm}
\usepackage[normalem]{ulem}
\usepackage{braket} 
\usepackage{amsmath}
\usepackage{cancel}
\usepackage{slashed}
\usepackage{multicol}
\usepackage[capitalise]{cleveref}
\usepackage{mathtools} 
\usepackage{bbm} 

\counterwithin*{equation}{section}
\allowdisplaybreaks

\newcommand{\nnl}{\nonumber \\}

\newcommand{\cL}{\mathcal{L}}
\newcommand{\cO}{\mathcal{O}}

\newcommand{\cM}{{\mathcal M}}

\newcommand{\re}{{\mathrm{Re}} \,}
\newcommand{\im}{{\mathrm{Im}} \,}

\newcommand{\hc}{\rm h.c.}

\usepackage{etoolbox,siunitx}
\robustify\bfseries

\usepackage{subcaption}

\begin{document}

\title{
Covariant approach to allowed beta decay
}

\author[a]{Sergio Cruz-Alzaga}
\author[b]{and Adam Falkowski}

\affiliation[a]{Instituto de F\'isica Corpuscular (IFIC), CSIC‐UV, Spain and Departament de F\'isica Te\`orica, UV, Spain}
\affiliation[b]{Universit\'{e} Paris-Saclay, CNRS/IN2P3, IJCLab, 91405 Orsay, France}

\emailAdd{
sergio.delacruz.alzaga@ific.uv.es,
adam.falkowski@ijclab.in2p3.fr
}

\abstract{

We propose a Lorentz-covariant on-shell formalism for allowed nuclear beta decay. 
Nuclear matrix elements are organized in a systematic effective field theory expansion in the recoil momentum, with the dependence on nuclear spin and polarization encoded in little-group-covariant structures. 
In this language,  we construct the most general standard and exotic currents for allowed transitions through first recoil order. 
We then derive the differential decay distribution encoded in a set of Lorentz-invariant form factors. 
Specializing to the rest frame of the parent nucleus, we relate our parameters to those used in the traditional beta-decay literature and in nucleon- and quark-level effective field theories.
The formalism provides a natural framework for extensions to higher recoil orders, radiative corrections, and forbidden transitions.
}
\maketitle

\section{Introduction} 
\label{sec:intr}

Nuclear beta transitions play a central role in our quest for understanding fundamental interactions. 
First of all, they give us a unique insight into strongly coupled Quantum Chromodynamics (QCD) phenomena. 
In addition, superallowed $0^+ \to 0^+$ transitions provide the best probe of the $V_{ud}$ entry of the Cabibbo-Kobayashi-Maskawa (CKM) matrix of the Standard Model (SM), with additional information offered also by neutron and mirror decays \cite{Gonzalez-Alonso:2018omy,Hardy:2020qwl,Falkowski:2020pma}. 
Beta decays also allow for phenomenological determinations of the axial nucleon charge $g_A$~\cite{Abele:2008zz,Falkowski:2020pma,Dubbers:2021wqv}. 
Beyond the SM (BSM), they probe dimension-6 operators representing leading deformations of the SM~\cite{Herczeg:2001vk,Gonzalez-Alonso:2018omy,Falkowski:2020pma}. 
The current sensitivity is to operators suppressed by a scale of order $10$~TeV~\cite{Gonzalez-Alonso:2018omy}, above the direct reach of the Large Hadron Collider and comparable to its indirect reach via Drell-Yan production~\cite{Cirigliano:2012ab,deBlas:2013qqa,Falkowski:2017pss,Dawson:2021ofa}. 

The theoretical description of beta transitions has a long history starting almost a century ago~\cite{Fermi:1934hr}, with the main ingredients worked out by the end of the 1950s~\cite{Lee:1956qn,Jackson:1957zz,Weinberg:1958ut}.
Quite naturally, the commonly used formalism and nomenclature contain elements that may seem opaque from today's perspective.
It may be advantageous to develop a language that more readily connects to recent advances in quantum field theory. 
This is the main purpose of this paper. 

In this paper we propose a Lorentz-covariant on-shell framework for nuclear beta transitions. 
An important ingredient is the spinor-helicity variables for massive particles developed in Ref.~\cite{Arkani-Hamed:2017jhn}, which allow one to write down nuclear amplitudes that are manifestly covariant under the little group transformations. 
Our approach naturally accommodates frame independence, arbitrary nuclear spins, and a systematic recoil expansion, while treating standard and exotic currents on the same footing. 
Elements of this formalism have already been used in Refs.~\cite{Falkowski:2021vdg,Alviani:2024fby}, but this paper provides the first complete covariant treatment at subleading recoil order. 

Applied to allowed beta decay, the covariant framework nicely combines with a systematic effective field theory (EFT) expansion of the nuclear currents in the recoil momentum.
At leading and first recoil order, the amplitudes can be parametrized in terms of a finite set of transition-dependent Wilson coefficients multiplying Lorentz- and little-group-covariant structures.
The standard (vector and axial) currents are incorporated together with exotic (scalar, pseudoscalar, and tensor) ones.  
Our approach bears some resemblance to that of Holstein in Ref.~\cite{Holstein:1974zf}, however the main difference is that not only momenta but also polarizations are treated fully covariantly. 
Namely, the dependence on the nuclear polarizations is encoded via spin vectors and higher-rank spin tensors, which are Lorentz- and little-group covariant functions of the massive spinors associated with the parent nucleus. 
This makes it possible to derive differential distributions without committing to the rest frame of the parent nucleus, and to recover the standard correlation coefficients only at the final stage.

For the standard currents, our Wilson coefficients can be simply mapped onto the traditional parametrization in the nuclear beta-decay literature~\cite{Holstein:1974zf}. 
A complete treatment of exotic currents through first recoil order is a relevant addition to the literature. 
Standard and exotic Wilson coefficients can also be related to the nucleon-level non-relativistic EFT (NR-EFT) approach of Ref.~\cite{Falkowski:2021vdg}. 
This connection provides a simple path to relate the Wilson coefficients to quark-level EFT parameters.
More generally, the proposed organization in terms of on-shell covariant structures provides a natural starting point for incorporating higher recoil orders, radiative corrections, and forbidden transitions.

This paper is organized as follows.
In \cref{sec:FORM} we introduce the spinor-helicity formalism for massive and massless particles and explain how it can be employed to construct beta decay amplitudes.
In \cref{sec:ABD} we specialize to allowed beta transitions. We derive the most general form of the vector, axial, scalar, pseudoscalar, and tensor currents through first recoil order in our framework. 
In \cref{sec:DIST} we use these amplitudes to obtain the fully differential decay distribution encoded in a set of form factors. 
Specializing to the rest frame of the parent nucleus, the latter are related to the familiar beta-decay correlation coefficients.
\cref{sec:DISC} is devoted to discussing the relation of our results to previous literature. 
In particular, we match our Wilson coefficients to parameters of traditional  nuclear-,  nucleon- and quark-level EFT descriptions of beta decay, before we conclude in \cref{sec:CON}.
Some technical details are postponed to the appendices.
In \cref{app:SI} we collect spinor identities used in the main text.
In \cref{app:EFT} we review the quark- and nucleon-level EFT descriptions.
In \cref{app:ERC} we collect the recoil-order contributions to form factors involving exotic currents.
Finally, in \cref{app:GT} we detail how the formalism is adapted to pure Gamow-Teller transitions, where the parent and daughter spins differ by one unit.

\section{Formalism} 
\label{sec:FORM}

In this section we briefly review how to associate spinor variables to momenta of massless and massive particles. 
The spinor formalism is then used to construct tree-level beta decay amplitudes that are Lorentz invariant and transform appropriately under the little group. 
Spinor notation and conventions reviewed in this section largely follow Ref.~\cite{Dreiner:2008tw}. 
Throughout we use the mostly minus metric signature, and the $\epsilon^{0123} = 1$ sign convention for the Levi-Civita tensor.

First of all, one defines the Lorentz vectors of sigma matrices: 
$\sigma^\mu = (1, \boldsymbol{\sigma})$, 
$\bar \sigma^\mu = (1, - \boldsymbol{\sigma})$,
with $\boldsymbol{\sigma} \equiv (\sigma^1,\sigma^2,\sigma^3)$ denoting the usual triplet of Pauli matrices.
A massless momentum
$p^\mu$, $p^2 =0$, can  be associated with a pair of two-component commuting spinors 
$\lambda_{\alpha}$, 
$\tilde \lambda_{\dot \alpha}$,
$\alpha=1,2$, via the relation 
$p \sigma \equiv 
p^\mu  [\sigma_\mu]_{\alpha \dot \beta} = \lambda_{\alpha} \tilde \lambda_{\dot \beta}$.\footnote{%
Throughout this paper, {\em momentum} always refers to four-momentum, unless otherwise noted.} 
The spinor indices can be raised  by an antisymmetric $\epsilon$ tensor: 
$\lambda^\alpha = \epsilon^{\alpha \beta} \lambda_\beta$,
$\tilde \lambda^{\dot \alpha} = \epsilon^{\dot \alpha \dot \beta} \tilde \lambda_{\dot \beta}$, where summation over repeated spinor indices is implicit.
One can then show that 
$[p \bar \sigma]^{\dot \alpha \beta} 
= \tilde \lambda^{\dot \alpha } \lambda^\beta$, 
and 
$(\lambda  \lambda) \equiv
 \lambda^\alpha  \lambda_\alpha  =0$, 
 $(\tilde \lambda \tilde \lambda) \equiv
 \tilde \lambda_{\dot \alpha} \tilde  \lambda^{\dot \alpha}  =0$.  
The latter implies that the spinors satisfy the Weyl equations: 
$0 = p \sigma \tilde \lambda 
\equiv  [p \sigma]_{\alpha \dot \beta} \tilde \lambda^{\dot \beta}$, 
and $0 = p\bar \sigma  \lambda 
\equiv  [p \bar \sigma]^{\dot \alpha \beta} \lambda_{\beta}$. 
The $U(1)$ subgroup of the little group associated with $p^\mu$ acts on the spinors as 
$\lambda\to t^{-1} \lambda$, 
$\tilde \lambda\to t \tilde \lambda$, as this transformation does not change the momentum $p^\mu$. 
We also define the antisymmetric combinations of sigma matrices: 
$\sigma^{\mu\nu} = {i \over 2} [ \sigma^\mu \bar \sigma^\nu - \sigma^\nu \bar \sigma^\mu ]$, 
and 
$\bar \sigma^{\mu\nu} = {i \over 2} [ \bar \sigma^\mu  \sigma^\nu - \bar \sigma^\nu  \sigma^\mu ]$. 

On the other hand, a massive momentum, $p^2=m^2$, can be associated with {\em two} pairs of two-component spinors 
$\chi_{\alpha}^K$, 
and $\tilde \chi_{\dot \alpha}^K$, $K=1,2$~\cite{Arkani-Hamed:2017jhn}. 
In addition to the spinor index $\alpha$, the spinors now have the $SU(2)$ little group index  $K$, so that little group acts as 
$\chi^K \to \chi^L \Omega_L{}^K$, 
$\tilde \chi_K \to \Omega^\dagger_K{}^L \tilde \chi_L$.
The little group indices can be raised and lowered in the analogous way as the spinor ones: $\chi_K = \epsilon_{KL} \chi^L$, 
$\tilde \chi^K = \epsilon^{KL} \tilde \chi_L$ (spinor indices are implicit from now on).
The defining relations connecting the massive spinors and momentum are 
$p \sigma =
\chi^K \tilde \chi_K$,
$\chi^{K} \chi_{L}
= \tilde \chi_{L} \tilde \chi^{K}
= m \delta^K_L$. 
From these, it follows that the massive spinors satisfy the Dirac equation, 
$p \bar \sigma \chi^K = m \tilde \chi^K$, 
$p \sigma \tilde \chi^K = m  \chi^K$.  

With the spinor formalism at hand, scattering amplitudes can be written in the form that makes transparent their little group covariance properties. 
Consider an amplitude 
$\cM[1^h_X 2_Y^{(J,J_3)}  \dots ]$ containing an incoming massless particle $X$ with momentum $p_1^\mu$ and helicity $h$, and an incoming massive particle $Y$ with momentum $p_2^\mu$, spin $J$, and polarization $J_3$ along some chosen quantization axis. 
The amplitude can be constructed out of momenta $p_1$, $p_2$, \dots and the corresponding spinors 
$\lambda_1$,  $\tilde \lambda_1$,  $\chi_2^{K_i}$ or $\tilde \chi_2^{K_i}$,  \dots. 
The general rules of quantum field theory imply little group covariance of scattering amplitudes.
That is to say, under the massless little group transformation 
$\lambda_1 \to t^{-1} \lambda_1$ and
$\tilde \lambda_1 \to t \tilde  \lambda_1$,
our amplitude must scale as 
$\cM[1_X^h 2_Y^{(J,J_3)} \dots ]  \to t^{2h}\cM[1_X^h  2_Y^{(J,J_3)} \dots ]$.
For massive particles, on the other hand, little group covariance can be realized if the amplitude is constructed out of exactly $2J$ spinors
$\chi_2^{K_i}$ or $\tilde \chi_2^{K_i}$,  such that it is a fully symmetric tensor in the 
$K_1 \dots K_{2J}$ indices. 
Each independent configuration of these indices represents a possible value of polarization: 
$J_3 = J$ corresponds to  
$(K_1,K_2,\dots,K_{2J})=(1,1,\dots,1)$, 
$J_3 = J-1$ corresponds up to normalization to $(1,1, \dots, 1, 2)$ or any permutation thereof,
$J_3 = J-2$ corresponds to $(1,1, \dots, 1, 2, 2)$, 
and so on. 
In other words, the amplitude can be represented as a symmetric tensor in the little group space: 
$\cM[1^h_X 2_Y^{K_1 \dots K_{2J} }  \dots ]$. 
We will work with all-incoming amplitudes where the momentum conservation imposes $\sum_i p_i = 0$.
Outgoing particles can be handled using crossing symmetry. 
For massless particles the crossing rule in our conventions is 
$\cM[ A  \to (-n)_X^h B ]   = 
s_h  \cM[ A n_{\bar X}^{-h} \to B]$, 
where $\bar X$ denotes the antiparticle of $X$, 
while $s_h = 1$ for bosons and $s_h = {\rm sgn} (h)$ for fermions.   
For massive particles crossing rule is  
$\cM[ A  \to (-n)_Y^{O_1 \dots O_{2J}} B ]   = (-)^{2J}  \cM[ A n_{\bar Y}{}_{O_1 \dots O_{2J}} \to B]$. 
The lower position of the little group indices encodes polarization flips (together with possible additional minus signs) due to crossing.  

We can now apply the formalism discussed above to beta decay. 
We consider the decay amplitudes 
\begin{align}  
\cM_{\beta^-} = \cM[1_{{\cal N}_1} \to  (-2)_{{\cal N}_2} (-3)_{e^-} (-4)_{\bar \nu}]
, \nnl 
\cM_{\beta^+} = \cM[1_{{\cal N}_1} \to  (-2)_{{\cal N}_2} (-3)_{e^+} (-4)_{\nu}]
, \end{align} 
 for a parent nucleus ${\cal N}_1$ with spin $J_1$ undergoing $\beta^-$ ($\beta^+$) decay into a daughter nucleus ${\cal N}_2$ with spin $J_2$, an electron $e^-$ (positron $e^+$), and an anti-neutrino $\bar \nu$ (neutrino $\nu$).
In the all-incoming convention, all these particles are described by momentum labels $p_1$, $p_2$, $p_3$, and $p_4$, respectively, satisfying $p_1^0 >0$,  $p_{2,3,4}^0 < 0$  (for outgoing particles, the momentum labels are the negatives of their physical momenta). 
Momentum conservation implies $p_1 + p_2 + p_3 + p_4 = 0$.    
We  also define the momentum transfer 
$q \equiv p_1 + p_2$. 
We will consider $\cM_{\beta^\mp}$ in the limit of unbroken isospin, in which case 
$p_1^2=p_2^2=m_{\cal N}^2$. 
Note that this implies $2 p_1 q = 2 p_2 q = q^2$.  
Moreover, $p_3^2=m_e^2$, while neutrinos are treated as massless, $p_4^2=0$. 
The corresponding spinor variables are  $\chi_1^{I}$, $\chi_{2}^O$, $\chi_{3}^K$, $\lambda_4$,  together with their twiddle counterparts. 
The spinors are related to the incoming momenta as 
\begin{align} 
p_1 \sigma =  \chi_1^{I} \tilde \chi_{1\, I}
, \quad 
p_2 \sigma = \chi_2^{O} \tilde \chi_{2\, O}
, \quad 
p_3 \sigma =  \chi_3^{K} \tilde \chi_{3\, K}
, \quad 
p_4 \sigma = \lambda_4 \tilde \lambda_4
, \end{align}
with an implicit sum over the little group indices of the massive spinors. 
In our conventions, the little group indices of the parent nucleus are up, and those of the daughter nucleus and the beta particle are down. 
Little group covariance dictates that
$\cM_{\beta^\mp}$ must be a tensor with  $2J_1+2J_2+1$ massive little group indices, 
$\cM_{\beta^\mp} \equiv [\cM_{\beta^\mp}]^{I_1 \dots I_{2J_1}}_{O_1 \dots O_{2J_2} \, K}$, 
which is fully symmetric separately in the 
$I_j$ and $O_j$ indices.
Moreover, it must transform as 
$\cM_{\beta^-} \to t^{-1} \cM_{\beta^-}$,  
$\cM_{\beta^+} \to t \cM_{\beta^+}$,  
under the little group transformations associated with the massless momentum $p_4^\mu$.

At tree level, beta decay amplitudes can be factorized into a product of nuclear and leptonic currents:\footnote{%
The relative overall sign between the $\beta^-$ and $\beta^+$ amplitudes follows from our conventions for defining fermionic initial and final states.}
\begin{align}
\label{eq:FORM_Mbeta-current} 
\cM_{\beta^-} = & 
 -\bigg \{  \big [  j_{V-}^\mu  + j_{A-}^\mu  \big ]
  (\tilde \chi_{3 \, K}  \bar   \sigma_\mu  \lambda_4 )  
  + \big [  j_{S-} + j_{P-}  \big ]
 (\chi_{3 \, K}   \lambda_4 )   
 +  \big [  j_{T-}^{\mu \nu}  \big ] 
 ( \chi_{3 \, K} \sigma_{\mu \nu} \lambda_4 )   
 \bigg \} 
 , \nnl 
 \cM_{\beta^+} = & 
  \big [  j_{V+}^\mu  + j_{A+}^\mu  \big ]
  (\tilde \chi_{3 \, K}  \bar   \sigma_\mu  \lambda_4 )^* 
+ \big [  j_{S+} + j_{P+}  \big ]
 (\chi_{3 \, K}   \lambda_4 )^* 
 + \big [  j_{T+}^{\mu \nu}  \big ] 
 ( \chi_{3 \, K} \sigma_{\mu \nu} \lambda_4 )^*  
,  \end{align}
where $j_{X\pm}$ represent the matrix elements of the vector, axial, scalar, pseudoscalar, and tensor current operators
$\hat j_{X \pm}(x)$ between the nuclear states: 
\begin{align} 
\label{eq:FORM_jX-definition} 
j_{X\pm}  = \langle  {\cal N}_2(-p_2)_{O_1 \dots O_{2J_2} }   | \hat j_{X\pm}(0) | {\cal N}_1(p_1)^{I_1 \dots I_{2J_1} }  \rangle 
.   \end{align}
The definition of the current operators in terms of the quark fields is given in \cref{eq:EFT_jX}. 
Implicitly, $j_{X\pm}\equiv [j_{X\pm}]^{I_1 \dots I_{2J_1}}_{O_1 \dots O_{2J_2}}$ is a  tensor with $2J_1 + 2J_2$ little group indices, where $I_i (O_i)$ encode the polarization $J_3$ of the parent (daughter) nucleus. 
In practice, these indices appear because $j_{X\pm}$ are functions of the massive spinors. 
Implicitly, all $2J_1$ spinors associated with the parent nucleus carry the upper little group index, $\chi_1^{I_i}$, 
while those associated with the daughter nucleus carry the lower index, $\chi_{2 \, O_i}$. 
Both upper and lower indices are implicitly fully symmetrized, such that $j_X$ is fully symmetric in both the upper and lower indices.\footnote{Mind that the upper and lower indices are not symmetrized between themselves.}

The discrete parity symmetry ${\cal P}$ acts on momenta as $p_i \to \tilde p_i \equiv  (p_i^0,-\boldsymbol{p}_i)$. 
Parity invariance of QCD implies that currents should transform under ${\cal P}$ as 
\begin{align}    
\label{eq:FORM_Pconstraint}    
j^{\mu_1 \dots \mu_n}_{X\pm}(p_1,p_2)  =   & 
\eta_1 \eta_2 (-)^X   (-)^{\mu_1} \dots   (-)^{\mu_n}    j^{\mu_1 \dots \mu_n}_{X\pm}(\tilde p_1,\tilde p_2)
,   \end{align} 
where  $\eta_i$ are intrinsic parities of the nuclei,
$(-)^\mu$ is defined as 
$(-)^0 \equiv 1$, $(-)^k \equiv -1$, 
and moreover we define 
$ (-)^V = (-)^S = (-)^T = 1$,  $(-)^A = (-)^P = -1$.
In the next section we will build currents out of massive spinors, ensuring that the parity constraint in \cref{eq:FORM_Pconstraint} is satisfied. 
The parity properties of spinor bilinears are collected in \cref{tab:CON_parity}.

Furthermore, if time reversal symmetry ${\cal T}$ is unbroken, the currents should transform under it (more precisely, under the product of ${\cal P T}$) as\footnote{%
Note that the parent and daughter nuclei are either both bosons or both fermions, thus $(-)^{2J_1} = (-)^{2J_2}$.}
\begin{align} 
\label{eq:FORM_Tconstraint} 
 [j^{\mu_1 \dots \mu_n}_{X\pm}(p_1,p_2)]^{I_1 \dots I_{2J_1}}_ {O_1 \dots O_{2J_2}}   =  & 
 - (-)^X (-)^{2J_1} \big \{ [j^{\mu_1 \dots \mu_n}_{X\pm}(-p_1,-p_2)]_{I_1 \dots I_{2J_1}}^{O_1 \dots O_{2J_2}} \big\}^*
. \end{align} 
The ${\cal PT}$ transformation properties of spinor bilinears are collected in \cref{tab:CON_parity}. 

Finally, strong dynamics in the isospin limit respects $G$-parity~\cite{Gell-Mann:1962yej}, which is a combination of charge conjugation with an isospin rotation.   
For $J_1 = J_2$ and transitions within an isospin multiplet, we can classify the currents into first- and second-class currents according to their $G$-parity transformation properties, with the latter expected to be suppressed~\cite{Weinberg:1958ut}.  
In our notation, currents with the first-class $G$-parity assignment satisfy the constraint 
   \begin{align}   
   \label{eq:FORM_Gconstraint}
j_{X\pm}^{\mu_1 \dots \mu_n}(p_1,p_2) 
= & -  (-)^X  (-)^{2J_1}  j_{X\pm}^{\mu_1 \dots \mu_n}(p_2,p_1), \qquad & X = V,A,T 
, \nnl
j_{X\pm}(p_1,p_2) 
= &  (-)^X  (-)^{2J_1}  j_{X\pm}(p_2,p_1), \qquad & X = S,P
 .    \end{align}

\begin{table}[h]
\centering
 \begin{tabular}{|c|c|c|} 
 \hline 
Bilinear  & ${\cal P}$ & ${\cal P T}$ 
\\ \hline 
 $\chi_2 \chi_1 \pm  \tilde \chi_2 \tilde \chi_1$
 &  
$\pm  $
& 
$\mp $ 
\\
$\chi_2  \sigma^\mu \tilde  \chi_1  \pm \tilde \chi_2  \bar \sigma^\mu \chi_1$
& 
$\pm (-)^\mu$
& 
$\pm  $
\\
$\chi_2 \sigma^{\mu \nu} \chi_1  \pm   \tilde \chi_2 \bar \sigma^{\mu \nu}   \tilde \chi_1$
 &  
$\pm (-)^\mu (-)^\nu$ 
& 
$\pm  $ 
\\ \hline 
$\chi_1 \sigma^\mu \tilde  \chi_1$
& 
$- (-)^\mu$
& 
$+$
\\
$\chi_1 \sigma^{\mu \nu} \chi_1  \pm   \tilde \chi_1 \bar \sigma^{\mu \nu}   \tilde \chi_1$
 &  
$\mp (-)^\mu (-)^\nu$ 
& 
$\pm $ 
\\ \hline 
\end{tabular}
 \caption{
Transformations of spinor bilinears under parity ${\cal P}$ acting as $p_i \to \tilde p_i \equiv  (p_i^0,-\boldsymbol{p}_i)$, 
and under ${\cal P T}$ acting as 
${\cal PT} F(p_1,p_2) \equiv   F(-p_1,-p_2)^*$. 
 }
  \label{tab:CON_parity}
\end{table}

So far, the discussion has been valid for arbitrary spins of the parent and daughter nuclei. 
In the remainder of this paper we restrict our discussion to {\em allowed} beta decay~\cite{Severijns:2006dr}. 
The following section covers the case of mixed Fermi-Gamow-Teller ($J_1 = J_2 > 0$, $\eta_1 = \eta_2$). 
The case of pure Fermi transitions ($J_1=J_2=0$, $\eta_1 = \eta_2$) can be trivially obtained from the mixed case,  while pure Gamow-Teller transitions ($J_1 = J_2 \pm 1$, $\eta_1 = \eta_2$)  will be discussed in \cref{app:GT}.

\section{Covariant approach to allowed beta decay} 
\label{sec:ABD}

In this section we study the mixed Fermi-Gamow-Teller allowed transitions in the on-shell formalism laid out in the previous section.  
That is to say,  we restrict the discussion to the specific case of  equal nuclear spins and parities: $J_1 = J_2 \equiv J$, $\eta_1 = \eta_2$.
We expand the currents in powers of the recoil momentum $q = p_1+p_2$.
In this power counting we also take into account that 
$\chi_2 \chi_1 - \tilde \chi_2  \tilde \chi_1 \sim q$, 
see \cref{eq:SI_RecoilExpansion}. 
In this paper we truncate the expansion at linear order in $q$ (subleading order in recoil), but it is straightforward to continue to higher orders if needed.  
Our immediate goal is to write down the most general form allowed by symmetries of the currents $j_{X\pm}$ up to subleading order in recoil. 
These currents are then inserted in \cref{eq:FORM_Mbeta-current} to construct the beta decay amplitude.   

\begin{figure}[tb]
    \centering
    \includegraphics[width=0.5\linewidth]{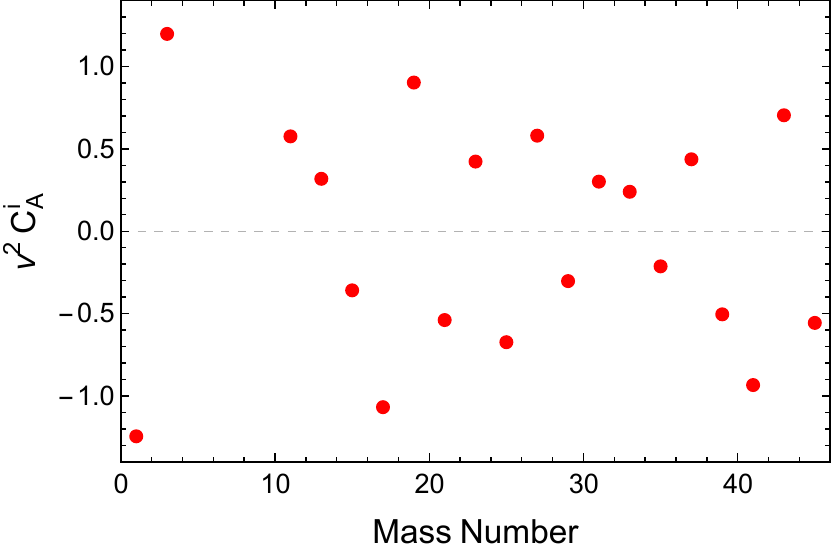}
    \caption{    \label{fig:Mirror-CAi} 
Phenomenological values  of $C_A^i$ for mirror transitions up to mass number $A=45$. 
To determine $C_A^i$ we use values of the mixing parameters collected in Ref.~\cite{Vanlangendonck:2022dud}. }  
\end{figure}

For the standard (vector and axial) currents contributing to $\beta^-$ decay the most general form up to subleading order in recoil and consistent with parity is the following:\footnote{%
The currents in \cref{eq:ABD_jVjA,eq:ABD_jSjPjT} are written in the limit of equal parent and daughter masses. 
This is a technicality to simplify derivations rather than a physical restriction, see the discussion below \cref{eq:ABD_jXplus-expanded}. 
} 
\begin{align}
\label{eq:ABD_jVjA}
j_{V-}^\mu = & 
M_F^-\bigg \{
\bigg ( C_V^i \big [ \chi_2  \sigma^\mu \tilde \chi_1 +  \tilde \chi_2  \bar \sigma^\mu \chi_1 \big ] 
  - i{ C_M^i -C_V^i  \over 2 m_{\cal N} }  
\big [ \chi_2  \sigma^{\mu \nu}  \chi_1 +  \tilde \chi_2  \bar \sigma^{\mu \nu} \tilde  \chi_1 \big ]  q_\nu   
\bigg )
{ \big [ \chi_2 \chi_1 +  \tilde \chi_2 \tilde \chi_1  \big ]^{2J-1} \over (2 m_{\cal N})^{2 J-1} }  
\nnl   &
+     \bigg (  C_{IS}^i   + {J- {1 \over 2}  \over 3 J} C_Q^i \bigg ) 
q^\mu { \big [ \chi_2 \chi_1 +  \tilde \chi_2 \tilde \chi_1  \big ]^{2J} \over (2 m_{\cal N})^{2 J} }
\nnl & 
+    {J- {1 \over 2}   \over J} C_Q^i  
\big [ \chi_2  \sigma^\mu \tilde \chi_1 -   \tilde \chi_2  \bar \sigma^\mu \chi_1 \big ] 
\big [ \chi_2  \sigma^\nu \tilde \chi_1 -   \tilde \chi_2  \bar \sigma^\nu \chi_1 \big ]  q_\nu 
{ \big [ \chi_2 \chi_1 +  \tilde \chi_2 \tilde \chi_1  \big ]^{2J-2} \over (2 m_{\cal N})^{2 J} } 
\bigg \}  
  + \cO(q^2)
,\nnl 
j_{A-}^\mu = & 
M_F^- \bigg \{   C_A^i  \big [ \chi_2  \sigma^\mu \tilde \chi_1 -  \tilde \chi_2  \bar \sigma^\mu \chi_1 \big ]  
 +   i { C_{IT}^i \over 2 m_{\cal N} }   
 \big [ \chi_2  \sigma^{\mu \nu}  \chi_1 -  \tilde \chi_2  \bar \sigma^{\mu \nu} \tilde  \chi_1 \big ]  q_\nu    
\bigg \}
{ \big [ \chi_2 \chi_1 +  \tilde \chi_2 \tilde \chi_1  \big ]^{2 J-1} \over (2 m_{\cal N})^{2 J-1} }  
\nnl & 
+ \cO(q^2) 
. \end{align}
In writing the above, we assumed $J \geq 1/2$; we comment on the $J=0$ case below \cref{eq:ABD_jXminus-expanded}. 
Here, $M_F^\mp$ is the Fermi matrix element which, 
for $\beta^\mp$ transitions within the same isospin multiplet, is given by 
$M_F^\mp = \delta^T_{T'}\delta^{T_3}_{T_3'\mp 1}
\sqrt{T(T+1)-T_3(T_3\pm 1)}$,
where $(T,T_3)$ and $(T',T_3')$ 
are the isospin quantum numbers of the parent and daughter nuclei. 
At leading order in recoil only two Wilson coefficients arise: $C_V^i$ (vector) and $C_A^i$ (axial), while at subleading order we have in addition $C_M^i$ (weak magnetism), $C_{IS}^i$ (induced scalar), $C_{IT}^i$ (induced tensor), and $C_Q^i$ (quadrupole, which exists for $J \geq 1$ only).  
The Wilson coefficients $C_X^i$ are a priori transition-dependent, which we stress by including the label $i$.  
However, one can show that, in the limit of unbroken isospin, $C_V^i$ is universal for all allowed transitions.
Furthermore, for the so-called mirror transitions~\cite{Severijns:2008ep,Severijns:2021lsp,Vanlangendonck:2022dud}, isospin conservation relates $C_M^i$ to the magnetic moments
$\mu_{{\cal N}_1}$ and $\mu_{{\cal N}_2}$ of the parent and daughter nuclei~\cite{Gell-Mann:1958sai,Calaprice:1976jbi}: 
${C_M^i \over C_V^i} = 
A {\mu_+ - \mu_-\over \mu_N}$, where $A$ is the mass number for the transition, 
the nuclear magneton is defined as 
$\mu_N = {e \over 2 m_p}$,
and $\mu_+$ and $\mu_-$ denote the magnetic moments of the nuclei with larger and smaller isospin quantum number $T_3$, respectively. 
Furthermore, one can verify that the terms multiplied by 
$C_{IS}^i$, $C_Q^i$, and $C_{IT}^i$  do not satisfy the constraint in \cref{eq:FORM_Gconstraint}. 
These are the second-class currents in Weinberg's classification~\cite{Weinberg:1958ut}, which only arise due to isospin breaking and are therefore expected to be suppressed. 
The induced scalar and quadrupole interactions also violate vector current conservation.
On the other hand, $C_A^i$ is not fixed by symmetries. 
While dimensional analysis suggests $|C_A^i| = {\cO(1) \over v^2}$, its precise value currently cannot be determined with enough accuracy from first principles, with the exception of the neutron decay.\footnote{%
Specifically, $C_A^{n \to p} = - {g_A V_{ud} \sqrt{1+\Delta_R^A } \over v^2} \approx -{1.25 \over v^2}$ in the SM limit,  
$v \equiv (\sqrt 2 G_F)^{-1/2} \approx 246.22$~GeV,
$G_F$ is the Fermi constant, the axial coupling of the neutron $g_A$ is known from lattice QCD with percent-level precision~\cite{FlavourLatticeAveragingGroupFLAG:2024oxs}, and the short-distance radiative corrections encoded in $\Delta_R^A$ are also known at the percent level~\cite{Cirigliano:2022hob}.}
In practice, $C_A^i$ is determined phenomenologically from the lifetime and/or correlation measurements.
See \cref{fig:Mirror-CAi} for example values of $C_A^i$ for a selection of mirror transitions.

For the exotic currents contributing to $\beta^-$ decay, for $J \geq 1/2$ the most general forms up to subleading order in recoil and consistent with parity are the following
\begin{align} 
\label{eq:ABD_jSjPjT} 
j_{S-}  = & 
 M_F^{-} C_S^i  { \big [ \chi_2 \chi_1 +  \tilde \chi_2 \tilde \chi_1  \big ]^{2J} \over (2 m_{\cal N})^{2J-1} } 
 + \cO(q^2) 
\nnl 
j_{P-}  = & 
 M_F^{-} C_P^i  \big [ \chi_2 \chi_1 -  \tilde \chi_2 \tilde \chi_1  \big ] 
 { \big [ \chi_2 \chi_1 +  \tilde \chi_2 \tilde \chi_1  \big ]^{2J-1} \over (2 m_{\cal N})^{2J-1} } 
 + \cO(q^2) 
 \nnl 
j_{T-}^{\mu \nu}   = & 
 {M_F^{-} \over 2 }  \bigg \{   
\bigg [  C_T^i  \big [  \chi_2  \sigma^{\mu \nu}  \chi_1 +  \tilde \chi_2  \bar \sigma^{\mu \nu} \tilde  \chi_1 \big ]   
 + i {  C_{T(1)}^i  - C_T^i    +  { (J - 1/2) C_{TQ}^i \over 3  J }  \over m_{\cal N} }   
 \big [  \chi_2  \sigma^{[\mu} \tilde  \chi_1 +  \tilde \chi_2  \bar \sigma^{[\mu}   \chi_1 \big ]  q^{\nu]} 
  \nnl  &
   +    { C_{T(2)}^i  \over 2 m_{\cal N} }\epsilon^{\mu\nu\alpha \beta}  
q_\alpha \big [  \chi_2  \sigma_\beta \tilde  \chi_1 - \tilde \chi_2  \bar \sigma_\beta   \chi_1 \big ]  
  \bigg ]     { \big [ \chi_2 \chi_1 +  \tilde \chi_2 \tilde \chi_1  \big ]^{2J-1} \over (2 m_{\cal N})^{2J-1} } 
  \nnl & 
 -   i  {(J-1/2) C_{TQ}^i \over J m_{\cal N}^2 }  
 \big [ \chi_2 \chi_1  - \tilde \chi_2 \tilde \chi_1  \big ] 
\big [ \chi_2 \sigma^{[\mu} \tilde \chi_1 -  \tilde \chi_2 \bar \sigma^{[\mu} \chi_1 \big ] p_1^{\nu]}  
{   \big [ \chi_2 \chi_1  + \tilde \chi_2 \tilde \chi_1  \big ]^{2J-2}   \over (2 m_{\cal N})^{2J-2} }  
  \bigg \}  
   + \cO(q^2)  
   , \end{align}   
where $X^{[\mu} Y^{\nu]} \equiv (X^{\mu} Y^{\nu}-X^{\nu} Y^{\mu})/2$. 
At leading order in recoil two more Wilson coefficients arise, $C_S^i$ (scalar) and $C_T^i$ (tensor). At subleading order we also have $C_P^i$ (pseudoscalar),  $C_{T(n)}^i$ with $n=1,2$ (subleading tensor), 
and $C_{TQ}^i$ (quadrupole tensor). 
One can show that $C_P^i$ and $C_{T(2)}^i$ do not satisfy the constraint in \cref{eq:FORM_Gconstraint}. 

For the standard and exotic currents in \cref{eq:ABD_jVjA,eq:ABD_jSjPjT},  the parity symmetry constraint in \cref{eq:FORM_Pconstraint} is satisfied, given the action of ${\cal P}$  on spinor bilinears collected in \cref{tab:CON_parity}.
Furthermore, the time reversal constraint  in \cref{eq:FORM_Tconstraint} is satisfied  provided the Wilson coefficients $C_X^i$ are real.

Using the recoil expansion of the spinor bilinears in \cref{eq:SI_RecoilExpansion}, we can rewrite the currents up to $\cO(q^2)$ corrections as 
  \begin{align}
\label{eq:ABD_jXminus-expanded}
j_{V-}^\mu   = & 
 M_F^- \bigg \{ 
2 C_V^i   p_1^\mu  \mathbbm{1} 
- \big [ C_V^i -   C_{IS}^i \big ] q^\mu  
\mathbbm{1}    
+  i{C_M^i    \over J m_{\cal N} }  
\epsilon^{\mu \nu \alpha \beta} 
q_\nu   p_1{}_\alpha  S_\beta  
 +{C_Q^i \over J^2  }   q_\nu  S^{\mu \nu}  
  \bigg \} 
  ,\nnl 
  j_{A-}^\mu   = & 
 M_F^- \bigg \{ 
 {2 m_{\cal N}  C_A^i  \over J}   S^\mu  
+ { C_A^i   +  C_{IT}^i  \over J m_{\cal N}  }    (q S)   p_1^\mu 
  \bigg \} 
, \nnl 
j_{S-}  =  & 
 M_F^{-} \bigg \{ 2 m_{\cal N} C_S^i \mathbbm{1} \bigg \}  
, \nnl 
j_{P-}  =  & 
  M_F^{-} \bigg \{   { C_P^i   \over J} (q S)  \bigg \}
  , \nnl 
j_{T-}^{\mu \nu}   =  & 
  i  M_F^{-}  \bigg \{   
 -  {    C_T^i   \over J }   \big [  p_1^\mu  S^\nu - p_1^\nu  S^\mu \big ]  
+  {C_{T(1)}^i \over 2 m_{\cal N} }  
\big[ p_1^\mu  q^\nu -  p_1^\nu  q^\mu    \big ]   \mathbbm{1} 
  +  {C_T^i  + C_{T(2)}^i  \over 2 J}    \big [ q^\mu S^\nu   - q^\nu S^\mu \big ]
\nnl & 
+ {C_{TQ}^i \over 2 m_{\cal N} J^2} \big [ p_1^{\mu}  S^{\nu \alpha} -  p_1^{\nu}  S^{\mu \alpha}   \big ]  q_\alpha  
  \bigg \}     
. \end{align} 
Above, 
$\mathbbm{1} \equiv  \delta^{I_1}_{O_1} \dots \delta^{I_{2J}}_{O_{2J}}$, 
with implicit full symmetrization of, independently, upper and lower indices. 
We also introduced the spin (pseudo)vector $S^\mu$~\cite{Holstein:2008sx} and the symmetric two-index spin tensor $S^{\mu \nu}$ defined in 
\cref{eq:SI_spinVector,eq:SI_spinTensor}. 
The spin vector encodes the information about polarization of the parent nucleus. 
It is transverse, that is to say $p_1^\mu S_\mu = 0$. 
The spin tensor  is symmetric in the Lorentz indices,  
$S^{\mu \nu} = S^{\nu \mu}$, 
transverse, $p_1^\mu S_{\mu \nu} = 0$, and traceless, 
$S^\mu_\mu = 0$. 
For $J=0$ (pure Fermi decays),  the spin-dependent structures $S^\mu$ and $S^{\mu \nu}$ are absent and only the $J$-independent terms are retained. 
For $J=1/2$, one should set $C_Q^i =  C_{TQ}^i = 0$. 
\cref{eq:ABD_jXminus-expanded} is an elegant and compact covariant expression for the currents, where the massive spinors are concealed within $S^\mu$ and $S^{\mu \nu}$. 
This form is particularly convenient for calculating differential distributions in the next section.  
 
We can formally relate $\beta^-$ and $\beta^+$ decay amplitudes by exploiting crossing symmetry and real analyticity of the amplitudes: 
  \begin{align}   
\cM_{\beta^+} = \cM_{\beta^-}^*|_{p_1 \leftrightarrow -p_2, p_3 \to -p_3, p_4 \to -p_4 }
    .\end{align} 
Using this, we can derive the $\beta^+$ currents from the \cref{eq:ABD_jXminus-expanded} ones: 
\begin{align}
   \label{eq:ABD_jXplus-expanded} 
j_{V+}^\mu  = & 
 M_F^+ \bigg \{ 
2 \bar C_V^i   p_1^\mu  \mathbbm{1} 
-   \big [ \bar C_V^i +  \bar C_{IS}^i \big ] q^\mu  
\mathbbm{1}    
+   i{\bar C_M^i    \over J m_{\cal N} }  
\epsilon^{\mu \nu \alpha \beta} 
q^\nu   p_1^\alpha  S^\beta  
-  {\bar C_Q^i \over J^2 }   q_\nu  S^{\mu \nu}  
  \bigg \}  
  ,\nnl  
  j_{A+}^\mu   = & 
   M_F^+ \bigg \{ 
 {2 m_{\cal N} \bar C_A^i  \over J}  S^\mu  
+  { \bar C_A^i   -  \bar  C_{IT}^i  \over J m_{\cal N}  }    (q S)   p_1^\mu 
  \bigg \} 
, \nnl 
j_{S+}  =  &   
 M_F^+  \bigg \{ - 2 m_{\cal N} \bar C_S^i \mathbbm{1} \bigg \}   
, \nnl 
j_{P+}  =  & 
 M_F^+ \bigg \{  { \bar C_P^i   \over J} (q S) \bigg \} 
  , \nnl 
j_{T+}^{\mu \nu}   =  & 
    i  M_F^+  \bigg \{   
 - { \bar C_T^i   \over J }   \big [  p_1^\mu  S^\nu - p_1^\nu  S^\mu \big ]  
-  {\bar C_{T(1)}^i \over 2 m_{\cal N} }  
\big[ p_1^\mu  q^\nu -  p_1^\nu  q^\mu    \big ]   \mathbbm{1} 
 +   { \bar C_T^i  - \bar C_{T(2)}^i  \over 2 J}    \big [ q^\mu S^\nu   - q^\nu S^\mu \big ] 
 \nnl & 
- {\bar C_{TQ}^i \over 2 m_{\cal N} J^2} \big [ p_1^{\mu}  S^{\nu \alpha} -  p_1^{\nu}  S^{\mu \alpha}   \big ]  q_\alpha  
  \bigg \}  
,   \end{align} 
where $\bar C_X^i \equiv (C_X^i)^*$. 

The parametrization of the currents in \cref{eq:ABD_jXminus-expanded,eq:ABD_jXplus-expanded} is the starting point for obtaining differential distributions for beta decay in the next section.
Before we move on, however, let us comment on one important technicality.
As mentioned earlier, the derivation of the currents in \cref{eq:ABD_jXminus-expanded}  from the spinorial currents in \cref{eq:ABD_jVjA,eq:ABD_jSjPjT} assumes equal masses of the parent and daughter nuclei.
Nevertheless, from now on we will consider that \cref{eq:ABD_jXminus-expanded,eq:ABD_jXplus-expanded} remain valid as they stand when 
$\Delta \equiv m_{{\cal N}}-   m_{{\cal N}_2} > 0$ 
($m_{{\cal N}} \equiv m_{{\cal N}_1}$).
This in fact {\em defines} our Wilson coefficients outside the isospin-symmetric mass limit.
Concretely, we could go back to  \cref{eq:ABD_jVjA,eq:ABD_jSjPjT}, redefining $C_X^i \to \hat C_X^i$, and repeating all the derivations for $\Delta > 0$. 
Doing that we would still arrive at \cref{eq:ABD_jXminus-expanded}, with 
$C_X^i = \hat C_X^i + \Delta \sum_Y M_{XY} \hat C_Y^i$, with some coefficients $M_{XY}$.
Restoring $\Delta >0$ is of course necessary for the sake of calculating differential distributions in \cref{sec:DIST}, as the final-state  phase space vanishes in the $\Delta \to 0$ limit. 
In our power counting we will treat the nuclear mass difference $\Delta$ as being of the same order as the recoil momentum,  $\Delta \sim \cO(q)$.

\section{Differential distributions} 
\label{sec:DIST}

Having provided in the previous section a Lorentz-covariant description of beta decay amplitudes, we move to discussing a Lorentz-invariant description of the differential decay width. 
We are interested in observables involving summation over polarizations of the beta particle and the daughter nucleus.  
We define the spin-summed amplitude squared: 
$|\overline{\cM}_{\beta}|^2 \equiv \sum_{O_1 \dots O_{2J_2}, K} | \cM_{\beta^\mp}|^2$, where $O_i$ and $K$ are the little group indices of the daughter nucleus and beta particle. 
For allowed decay, that amplitude squared can be parametrized as 
 \begin{align}   
 \label{eq:DIST_Msquared-parametrization}
|\overline{\cM}_{\beta}|^2  =  &   
   8 \kappa   
\bigg \{   (p_1 p_3) (p_1 p_4)    {\cal N } 
+ \bigg [ (p_1 p_3) (p_1 p_4)  
- m_{\cal N}^2  (p_3 p_4) \bigg ]  \aleph  
\nnl &
 - {   m_{\cal N } (S p_3)   \over J  }    \big [
 (p_1 p_4) {\cal A }    
 +  (p_3 p_4)  {\cal A } '  \big ] 
 -    { m_{\cal N }  (S p_4 )   \over J   } \big [   {\cal  B}   (p_1 p_3)  
 +  (p_3 p_4)  {\cal B } '  \big ] 
   \nnl &   
  + {S^{\mu \nu} \over J^2}   
   \bigg  [  2 m_{\cal N}^2 p_3^\mu p_4^\nu  {\cal C} 
  +    p_3^\mu p_3^\nu   (p_1 p_4) {\cal C}_{3}'   
   +    p_4^\mu p_4^\nu   (p_1 p_3) {\cal C}_{4}'    \bigg ] 
      \nnl &   
 -   {  m_{\cal N } \epsilon_{\mu \nu \alpha  \beta }  p_1^\mu  S^\nu p_3^\alpha  p_4^\beta 
   \over J   }  {\cal  D }   
+ {   \epsilon_{\mu \nu \alpha \beta} p_1^\mu  S^{\nu \rho} p_3^\alpha p_4^\beta  
\over J^2   }   
   \big [  p_3^\rho    {\cal D}_3' 
   +  p_4^\rho  {\cal D}_4'    \big ] 
\nnl &   
+ {2 m_{\cal N }  S^{\mu \nu \rho}  p_3^\mu p_4^\nu \over J^3  } 
\big [  {\cal E}_3'  p_3^\rho  + {\cal E}_4'  p_4^\rho  \big ]  
     \bigg \}  + \cO(m_{\cal N}^{-2} ) 
,    \end{align}  
where $\kappa$ is a numerical factor fixed to be 
$\kappa =  (M_F^\mp)^2$ for allowed decays with $J \equiv J_1 = J_2$, 
$\kappa= 2$ for pure Gamow-Teller decays with $J  = J_2 + 1$, 
and $\kappa = {2 (2 J + 3) \over 2 J + 1}$ 
 for pure Gamow-Teller decays with $J  = J_2 - 1$.   
\cref{eq:DIST_Msquared-parametrization} defines  the form factors 
$ {\cal N }$,  $ \aleph$,   
$ {\cal A }$,  $ {\cal A }'$,
$ {\cal B }$,  $ {\cal B }'$,
${\cal C}$,   ${\cal C}_i'$, 
$ {\cal  D }$,  ${\cal D}_i'$, ${\cal E}_i'$, 
which are bilinear in the Wilson coefficients $C_X^i$ and  have dimensions $[{\rm mass}]^{-4}$.  
They are in general  functions of the invariant $(p_1 p_3)$ (or equivalently, of  $E_e \equiv - (p_1 p_3)/m_{\cal N}$) as well as parameters such as the electron mass $m_e$, the parent nucleus mass $m_{\cal N}$,   and the nuclear mass difference 
$\Delta$
(or equivalently, the maximum relativistic energy of the beta particle in the rest frame defined as  $E_0 =  \Delta 
+ {m_e^2 - \Delta^2  \over  2 m_{\cal N} }$). 
We assume their eventual dependence on $(p_1 p_4)$ has been eliminated using the identity 
$(p_1 p_4)   =    m_{\cal N} (E_e   - E_0) 
-  (p_3 p_4) $, which follows from momentum conservation and on-shell conditions.  
By definition, the form factors do not depend on $(p_3 p_4)$. 
The form factors $ {\cal N }$,  $ \aleph$, $ {\cal A }$,  $ {\cal B }$,  ${\cal C}$ , $ {\cal  D }$,   
arise at leading order in recoil,  and they are closely related to the usual correlation coefficients 
 $a$, $A$, $B$, $c$ and $D$ and the Fierz term $b$ defined in Ref.~\cite{Jackson:1957zz}, see below in \cref{eq:DIST_correlations}.  
These form factors in general receive also recoil-order corrections, 
while the primed ones:  $ {\cal A }'$,   $ {\cal B }'$, ${\cal C}_i'$,   ${\cal D}_i'$, ${\cal E}_i'$, arise only at the recoil order. 
The form factors ${\cal  D }$ and ${\cal D}_i'$ are $T$-odd, while the remaining ones are $T$-even.
Beyond subleading order in recoil (that is at  $\cO(m_{\cal N}^{-2})$) additional tensor structures are possible, but we will not discuss them in this paper. 
The last line of \cref{eq:DIST_Msquared-parametrization} features the 3-index spin tensors defined in  
\cref{eq:SI_spinTensor}. 
Implicitly, 
$S^\mu$ is kept only for $J \geq 1/2$, 
$S^{\mu \nu}$ is kept for $J \geq 1$, 
and $S^{\mu \nu\rho}$ is kept for $J \geq 3/2$.

A straightforward if tedious calculation allows one to determine the dependence of the form factors on the Wilson coefficients $C_X^i$. 
Below we present the complete leading order result, while at the first recoil order we only give the dependence on  the Wilson coefficients entering the standard currents: 
\begin{align}
\label{eq:DIST_formfactors} 
{\cal N} = & \re \bigg \{  
|C_V^i|^2  + |C_S^i|^2 
\pm 2 {m_e \over E_e} C_V^i \bar C_S^i 
+  \lambda_0 \bigg [ |C_A^i|^2   + |C_T^i|^2 
\pm 2 {m_e \over E_e} C_A^i \bar C_T^i  \bigg ] 
\nnl & 
- {m_e^2 \over m_{\cal N} E_e} 
C_V^i (\bar C_V^i  \mp \bar C_{IS}^i) 
+ {\lambda_0 \over 3}  \bigg [ 
\mp 2 { 2 E_e - E_0 - {m_e^2 \over E_e}  \over m_{\cal N} }  C_A^i \bar C_M^i   
\pm { E_0 - {m_e^2 \over E_e}  \over  m_{\cal N} } 
C_A^i ( \bar C_{IT}^i  \pm  \alpha_A \bar C_A^i  )  \bigg ]  
\bigg \} 
, \nnl
\aleph = & \re \bigg \{   
|C_V^i|^2   - |C_S^i|^2 
 +  {\lambda_0  \over 3}  
 \bigg [ - |C_A^i|^2+  |C_T^i|^2  
\pm  {2  (2 E_e - E_0 )
\over   m_{\cal N} } 
C_A^i \bar C_M^i  
\pm { E_0 \over    m_{\cal N} }
 C_A^i ( \bar C_{IT}^i  \pm   \alpha_A \bar C_A^i  )   
 \bigg  ]  
\bigg \} 
, \nnl
{\cal A}  = & \re \bigg \{   
\mp  \lambda_1 \big [ |C_A^i|^2  - |C_T^i|^2  \big ]
- 2 C_V^i \bar C_A^i + 2 C_S^i \bar C_T^i 
\mp  {E_0 \over m_{\cal N}}  C_V^i \bar C_M^i 
- {2 E_e \over m_{\cal N} } 
 C_V^i (\bar C_A^i \pm\bar C_{IT}^i) 
\nnl  &  
+{\lambda_1 \over 2}  \bigg [  
 {2 E_e - E_0 \over   m_{\cal N} } C_A^i \bar C_M^i  
-  {E_0 \over   m_{\cal N} }
 C_A^i ( \bar C_{IT}^i  \pm   \alpha_A \bar C_A^i  ) \bigg ] 
 \pm  \lambda_3  { 6 E_e + E_0  \over   10 m_{\cal N} }    C_A^i \bar  C_Q^i    
 \bigg \}  
, \nnl
{\cal B}  = &  \re \bigg \{  
\pm \lambda_1 \bigg [   |C_A^i|^2 +  |C_T^i|^2 
\pm 2 {m_e \over  E_e}  C_A^i \bar C_T 
\bigg ] 
  - 2 C_V^i \bar C_A^i  - 2 C_S^i \bar C_T^i 
\mp 2 {m_e \over E_e} C_V^i \bar C_T 
\mp 2 {m_e \over E_e} C_A^i \bar C_S 
\nnl & 
\pm   {E_0 -  {m_e^2 \over E_e}  \over m_{\cal N} }   C_V^i \bar C_M^i 
+ {2 (E_e - E_0)  \over m_{\cal N} } 
C_V^i (\bar C_A^i \pm \bar C_{IT}^i)  
+ {m_e^2 \over m_{\cal N} E_e} 
C_A^i (\bar C_V^i  \mp  \bar C_{IS}^i ) 
\nnl & 
+ \lambda_1 \bigg [ -   {2 E_e - E_0 - {m_e^2 \over E_e}  \over 2   m_{\cal N} }  C_A^i \bar C_M^i   
+  { E_0  - {m_e^2 \over E_e } \over  2  m_{\cal N} }  C_A^i ( \bar C_{IT}^i  \pm   \alpha_A \bar C_A^i  )  
\bigg ] 
\pm {\lambda_3 \over 10 } { 7 E_0  -  6 E_e - {m_e^2 \over E_e}  \over m_{\cal N} }    C_A^i \bar  C_Q^i 
 \bigg \}  
, \nnl
{\cal C}  = &  
\re \bigg \{   
 \lambda_2 \big [ |C_A^i|^2 - |C_T^i|^2 \big ]  
 \pm { E_0  \over   2 m_{\cal N} }  C_V^i \bar  C_Q^i 
+ \lambda_2 \bigg [  
\mp {2 E_e - E_0 \over 2   m_{\cal N}  }  
C_A^i \bar C_M^i   
\pm  {E_0 \over 2   m_{\cal N}  } 
 C_A^i ( \bar C_{IT}^i  \pm   \alpha_A \bar C_A^i  )
 \bigg ] 
\nnl & 
 + {3 \lambda_4  (2 E_e - E_0) \over 2  m_{\cal N} } C_A^i \bar  C_Q^i  
 \bigg \}  
, \nnl
{\cal D}  = &  \im \bigg \{   
- 2 C_V^i \bar C_A^i  
+ 2 C_S^i \bar C_T^i 
\nnl & 
\pm  { 2 E_e - E_0 \over m_{\cal N}  }   
C_V^i \bar C_M^i  
- \lambda_1 \bigg [  
 {E_0 \over 2  m_{\cal N} }  C_A^i \bar C_M^i  
+   { 2 E_e - E_0 \over 2 m_{\cal N}  } 
 C_A^i \bar C_{IT}^i 
 \bigg ] 
 \pm  \lambda_3 {  E_0  \over 2 m_{\cal N} }    C_A^i \bar  C_Q^i 
\bigg \}  
, \end{align} 
\begin{align}
\label{eq:DIST_formfactors-primed} 
{\cal A}'  = &  \re \bigg \{ 
\mp C_V^i \bar C_M^i 
-  C_V^i (\bar C_A^i \pm \bar C_{IT}^i) 
-  {\lambda_1 \over   2  }  \bigg [  
 C_A^i \bar C_M^i 
+ C_A^i ( \bar C_{IT}^i  \pm   \alpha_A \bar C_A^i  )  
\bigg ] 
\mp {3 \lambda_3 \over 10 } C_A^i \bar  C_Q^i    
\bigg \} 
, \nnl 
{\cal B}'  = & 
 \re \bigg \{ 
 \pm C_V^i \bar C_M^i 
- C_V^i (\bar C_A^i \pm \bar C_{IT}^i) 
-  {\lambda_1 \over   2  }  \bigg [   
 C_A^i \bar C_M^i  
-  C_A^i ( \bar C_{IT}^i  \pm   \alpha_A \bar C_A^i  )
\bigg ] 
\mp { 3 \lambda_3 \over 10 } C_A^i \bar  C_Q^i   
\bigg \} 
, \nnl 
{\cal C}_{3}'  = & 
\re \bigg \{  
 \mp C_V^i \bar  C_Q^i  
\mp  \lambda_2 \bigg [  C_A^i \bar C_M^i  
+   C_A^i ( \bar C_{IT}^i  \pm   \alpha_A \bar C_A^i  ) \bigg ] 
+  3 \lambda_4  C_A^i \bar  C_Q^i   
\bigg \}  
, \nnl 
{\cal C}_{4}'  = & \re \bigg \{   
\mp  C_V^i \bar  C_Q^i    
 \pm \lambda_2 \bigg [  C_A^i \bar C_M^i 
- C_A^i ( \bar C_{IT}^i  \pm   \alpha_A \bar C_A^i  ) 
\bigg ] 
 -  3 \lambda_4  C_A^i \bar  C_Q^i    
 \bigg \}   
, \nnl 
 {\cal D}_3'  = & 
   \im  \bigg \{  
   \mp C_V^i \bar C_Q^i  
\pm    \lambda_2 \bigg [   C_A^i \bar C_M^i  
+  C_A^i \bar C_{IT}^i  
\bigg ]  
 +  \lambda_4 C_A^i \bar  C_Q^i  
 \bigg \} 
 , \nnl 
  {\cal D}_4'  = &  
     \im  \bigg \{  
     \mp   C_V^i \bar C_Q^i   
 \mp  \lambda_2 \bigg [   C_A^i \bar C_M^i 
 -   C_A^i \bar C_{IT}^i  \bigg ]   
    -   \lambda_4  C_A^i \bar  C_Q^i    
    \bigg \}  
 , \nnl  
{\cal E}_3' = & {\cal E}_4' =   \re \bigg \{ 
\mp  \lambda_5 C_A^i \bar  C_Q^i 
\bigg \}    
. \end{align}
Here, the upper (lower) signs refer to $\beta^-$ ($\beta^+$) decay. 
The spin-dependent group-theoretical coefficients 
$\lambda_i$ are collected in the second column of  \cref{tab:CORR_GroupTheoreticalLambda}. 
In this form, the results can also be used for pure Gamow-Teller decays, after setting $C_V^i = C_S^i = C_{IS}^i = C_{T(1)}^i = 0$ and with corresponding $\lambda_i$ and $\alpha_X$ coefficients given in the third and fourth columns of \cref{tab:CORR_GroupTheoreticalLambda}.\footnote{%
In \cref{tab:CORR_GroupTheoreticalLambda}, apparent singularities for some values of $J$  occur outside the range of $J$ for which the corresponding coefficients are used. 
For example, $\lambda_5$ is only used for $J \geq 3/2$, 
while $\lambda_{2,4}$ are only used for $J \geq 1$. 
}
See \cref{app:GT} for the conventions for currents and Wilson coefficients in the pure Gamow-Teller case.
For pure Fermi decays ($J=J_2 =0$) one should set instead $C_A^i = C_T^i = C_M^i = C_{IT}^i = C_P^i =  C_{T(2)}^i = C_Q^i =  C_{TQ}^i = 0$. 
The subleading recoil dependence of the form factors on the exotic Wilson coefficients is given in \cref{app:ERC}.

\begin{table}[th]
\begin{center} 
\begin{tabular}{|c|c|c|c|}
\hline 
Coefficient & $J= J_2$  & $J = J_2+1$  & $J = J_2 - 1$ 
\\ \hline 
$\lambda_0$ &  ${J+1 \over J}$  &  1 &  1 
\\ \hline 
$\lambda_1$ & ${1 \over J}$  & 1 &  $-{J \over J+1}$   
\\ \hline 
$\lambda_2$ & 1 &  $- {J \over 2 J - 1} $ &   $ - {J^2 \over (J+1)(2J+3) } $     
\\ \hline 
$\lambda_3$ & $  - {(2 J - 1) (2 J+3) \over 6 J^2 } $  &  1 &   ${J+2 \over J+1}$   
\\ \hline 
$\lambda_4$ &  $  {1 \over 2 J} $  &   $ { J  \over  2 J - 1 }   $ &  $  - {J(J+2) \over (J+1) (2 J + 3) }$  
\\ \hline 
$\lambda_5$ & 1 &  $ {2 J^2 \over  (J-1) (2J -1)  }$ &  ${2 J^2 \over (J+1) (2 J + 3) }  $ 
\\ \hline 
$\alpha_A$ & 1 &  0  & 2 
\\ \hline 
$\alpha_T$ & 1 & 0  & 0  
\\ \hline 
\end{tabular} 
\end{center} 
\caption{\label{tab:CORR_GroupTheoreticalLambda}
Group-theoretical factors appearing in the form factors and correlation coefficients for allowed beta decays. 
The columns refer to mixed Fermi-Gamow-Teller decays ($J \equiv J_1 = J_2$),  pure Gamow-Teller decays with $J = J_2 + 1$,  and those with $J = J_2 - 1$. }
\end{table}    

Given the amplitude squared, the differential decay width is given by 
$d \Gamma = {|\overline{\cM}_{\beta}|^2 \over 2 m_{\cal N} } d \Pi_3$, 
where the 3-body phase space element is 
$d \Pi_3 =  (2 \pi)^4 \delta^4(p_1 + p_2 + p_3 + p_4)  
\prod_{i=2}^4  {d^4 p_i \over (2 \pi)^3 }  \delta(p_i^2 - m_i^2) \theta(-p_i^0)$. 
In our formalism we can calculate the differential decay width in any reference frame. 
Traditionally, however, it is presented in the rest frame of the parent nucleus. 
In order to link to known results, we parametrize the rest frame momenta as 
$p_1 =  \big ( m_{\cal N}, \boldsymbol{0} \big)$, 
$p_3 =  -  \big ( E_e , \boldsymbol{k}_e \big)$, 
$p_4 = -  \big ( E_\nu , \boldsymbol{k}_\nu \big)$, 
where $E_e$ is the energy of the beta particle, 
$\boldsymbol{k}_i$ are 3-vectors describing spatial momenta of the beta particle and the neutrino whose magnitudes are fixed by on-shell conditions as 
$|\boldsymbol{k}_e| = \sqrt{E_e^2 -m_e^2}$, 
$|\boldsymbol{k}_\nu| = E_\nu$,  
while the neutrino energy is fixed by the kinematics as\footnote{%
Note that $\boldsymbol{k}_\nu/E_\nu$ is independent of neutrino energy, therefore the right-hand side of this equality does not depend on $E_\nu$. }
$E_\nu = { E_0  -  E_e  \over 
1 -  {  E_e E_\nu  - \boldsymbol{k}_e \boldsymbol{k}_\nu  \over   m_{\cal N} E_\nu }   }$. 
We also assume the parent nucleus state is prepared with a definite spin projection $J_3 \in [-J,J]$ along the spin quantization axis characterized by the unit vector 
$\boldsymbol{j}$. 
In that frame and for such a state, the spin vector is  $S =   \big ( 0 , \boldsymbol{J} \big )$, 
where the 3-vector $\boldsymbol{J} = J_3 \boldsymbol{j} $. 
Furthermore, the components of the  spin tensors take the values  
$S^{00} = S^{0k} = 0$,  
$S^{k l} =  \big [ \delta^{kl} - 3 \boldsymbol{j}^k \boldsymbol{j}^l  \big ]  {J (J+1) - 3 (\boldsymbol{J} \boldsymbol{j} )^2  \over 6} $, 
and $S^{000} = S^{0kl} = S^{00k} = 0$,
$S^{k l r } =
- { (\boldsymbol{J} \boldsymbol{j} )  
[ 5 (\boldsymbol{J} \boldsymbol{j} )^2 + 1  
- 3 J (J+1)  ]  \over 10 } \big \{ 
\delta^{kl}  \boldsymbol{j}^r   + 
 \delta^{lr}  \boldsymbol{j}^k +  \delta^{rk}  \boldsymbol{j}^l 
-  5 \boldsymbol{j}^k \boldsymbol{j}^l \boldsymbol{j}^r \big \} 
$.   
The phase space element in the rest frame can be written as  $ d \Pi_3 =    { d E_e  d \Omega_e d \Omega_\nu  \over 8  m_{\cal N}  (2\pi)^5  }
|\boldsymbol{k}_e|  (E_0 - E_e) 
\big (     1  -  {  E_e E_\nu   -  \boldsymbol{k}_e  \boldsymbol{k}_\nu  \over   m_{\cal N}  E_\nu}    \big )^{-2}$, where 
$d \Omega_i = d\cos \theta_i d \phi_i$ are the angular variables of the beta particle and neutrino momenta. 
Putting it all together, the differential distribution in the rest frame of the parent nucleus takes the form\footnote{%
Resummation of leading Coulomb interactions between the daughter nucleus of charge $Z$ and the beta particle amounts to multiplying the right-hand side by an overall Fermi function $F(Z,E_e)$~\cite{Fermi:1934hr}, see e.g.~\cite{Hayen:2017pwg}. 
}

 \begin{align}   &  
  \label{eq:CORR_distribution-restframe}   
 { d \Gamma \over  d E_e d \Omega_e       d \Omega_\nu  }=
 { \kappa  E_e |\boldsymbol{k}_e|  (E_0 - E_e)^2 \over     64 \pi^5  }
\bigg \{   \bigg ( 1   + {  3 E_e  \over   m_{\cal N} }    \bigg )  {\cal N } 
      - {3 (\boldsymbol{k}_e  \boldsymbol{k}_\nu)^2  \over   m_{\cal N}  E_e   E_\nu^2}   \aleph    
  +  {  (\boldsymbol{k}_e \boldsymbol{k}_\nu) \over  E_e E_\nu  } 
   \bigg [   \aleph      + {  3 E_e  \over   m_{\cal N} }   \bigg (   \aleph   -  {\cal N } \bigg )      
     \bigg ]  
\nnl &
 +  { (\boldsymbol{J}  \boldsymbol{k}_e )  \over J  E_e }    
 \bigg [   {\cal A }     + {  3 E_e  \over   m_{\cal N} }  {\cal A }   - {E_e \over  m_{\cal N} }  {\cal A } '
 + { (\boldsymbol{k}_e \boldsymbol{k}_\nu)  \over m_{\cal N} E_\nu }  \bigg ( {\cal A } '   - 3   {\cal A }  \bigg ) 
 \bigg  ] 
 \nnl & 
 +  {   (\boldsymbol{J}  \boldsymbol{k}_\nu )    \over J  E_\nu} 
 \bigg [   {\cal  B}      + {  3 E_e  \over   m_{\cal N} } {\cal  B}     - { E_\nu \over  m_{\cal N} }  {\cal B } '  
 +   { (\boldsymbol{k}_e \boldsymbol{k}_\nu) \over  m_{\cal N}  } 
 \bigg (  { {\cal B } '  \over E_e} - {3  {\cal  B}  \over E_\nu}  \bigg )    \bigg ] 
   \nnl &  
  +   {J (J+1) - 3 (\boldsymbol{J} \boldsymbol{j} )^2  \over 6 J^2}
  \big [ \delta^{kl} - 3 \boldsymbol{j}^k \boldsymbol{j}^l  \big ]    
   \bigg  [   { 2  \boldsymbol{k}_e^k  \boldsymbol{k}_\nu^l  \over E_e E_\nu }  
    \bigg (  1   + {  3 E_e  \over   m_{\cal N} }  
    - {3 (\boldsymbol{k}_e  \boldsymbol{k}_\nu)  \over   m_{\cal N}  E_\nu}     \bigg )  {\cal C} 
  -   {   \boldsymbol{k}_e^k  \boldsymbol{k}_e^l \over  m_{\cal N} E_e  } {\cal C}_{3}'   
   -  {  \boldsymbol{k}_\nu^k  \boldsymbol{k}_\nu^l  \over  m_{\cal N}  E_\nu} {\cal C}_{4}'    \bigg ] 
      \nnl &   
       +    {  \boldsymbol{J} \cdot (\boldsymbol{k_e} \times \boldsymbol{k_\nu})
   \over J  E_e E_\nu  }   \bigg  (  1   + {  3 E_e  \over   m_{\cal N} }  
      - {3 (\boldsymbol{k}_e  \boldsymbol{k}_\nu)  \over   m_{\cal N}  E_\nu}     \bigg )   {\cal  D }   
+ {J (J+1) - 3 (\boldsymbol{J} \boldsymbol{j} )^2  \over 2 J^2 }   {     
(\boldsymbol{k_e} \times \boldsymbol{k_\nu} ) \cdot   \boldsymbol{j}  
\over   m_{\cal N} E_e E_\nu     }   
   \big [ 
   (\boldsymbol{j} \boldsymbol{k}_e)    {\cal D}_3'  + (\boldsymbol{j}  \boldsymbol{k}_\nu) {\cal D}_4'    \big ] 
     \nnl & 
-  { (\boldsymbol{J} \boldsymbol{j} )  \big [ 5  (\boldsymbol{J} \boldsymbol{j} )^2 + 1  - 3 J (J+1) \big ] 
\over 5 J^3  } 
{  \boldsymbol{k}_e^k   \boldsymbol{k}_\nu^l   \over m_{\cal N }   E_e E_\nu } 
\big [   {\cal E}_3' \boldsymbol{k}_e^r + {\cal E}_4' \boldsymbol{k}_\nu^r   \big ]  
  \bigg [ 
 \delta^{kl} \boldsymbol{j}^r  + \delta^{kr} \boldsymbol{j}^l +  \delta^{lr} \boldsymbol{j}^k  -  5  \boldsymbol{j}^k \boldsymbol{j}^l \boldsymbol{j}^r \bigg ] 
      \bigg \} 
     \nnl &  
     + \cO(m_{\cal N}^{-2} )
.  \end{align}  
Given this, one can identify how the familiar correlation coefficients are related to the form factors:
 \begin{align}   
 \label{eq:DIST_correlations}
\xi  a =   &   \aleph      
+ {  3 E_e (\aleph   -  {\cal N }) \over   m_{\cal N} }  
, \nnl 
\xi A  =   &   {\cal A }     + {  3 E_e  {\cal A }   - E_e (  {\cal  B}  +  {\cal A } '  )  
 + {1   \over 3 } (E_0 - E_e)  {\cal B } '   
\over   m_{\cal N} }  
, \nnl
\xi  B   =   &   {\cal B }   + {  3 E_e {\cal  B}      - {(E_e^2 - m_e^2)\over E_e}{\cal A } 
- (E_0  - E_e)  {\cal B } '    +  {E_e^2 - m_e^2 \over 3   E_e}{\cal A } '   
   \over   m_{\cal N}  } 
   , \nnl
\xi  c   =   & 
 {2 J - 1 \over J}   \bigg (  1   + {  3 E_e  \over   m_{\cal N} }   \bigg ) {\cal C}    
   , \nnl
\xi  D   =   &  
    \bigg (  1   + {  3 E_e  \over   m_{\cal N} }   \bigg ) {\cal D} 
.  \end{align}   
Above, $\xi$ is an arbitrary normalization factor, whose role is to render the correlation coefficients dimensionless. 
Conventionally,  one takes  
$\xi \equiv  |C_V^i|^2 + |C_S^i|^2  + {J+1 \over J}  \big ( |C_A^i|^2 + |C_T^i|^2 \big )$. 
The beta–neutrino angular correlation coefficient (little $a$) is defined as the coefficient multiplying the 
${\boldsymbol{k}_e \boldsymbol{k}_\nu \over  E_e E_\nu  }$ structure in the curly bracket in \cref{eq:CORR_distribution-restframe}. 
Similarly, the $c$ and $D$ correlations are defined, respectively, 
as the coefficients of the 
$ {J (J+1) - 3 (\boldsymbol{J} \boldsymbol{j} )^2  
\over 3 J (2J-1) E_e E_\nu  }
  \big [ (\boldsymbol{k}_e  \boldsymbol{k}_\nu) 
  - 3 (\boldsymbol{k}_e \boldsymbol{j}) ( \boldsymbol{k}_\nu \boldsymbol{j})  \big ]$   
and ${\boldsymbol{J} (\boldsymbol{k_e} \times \boldsymbol{k_\nu})  \over J  E_e E_\nu  }$ structures. 
The beta asymmetry (capital $A$) is defined by  
${d \Gamma \over  d E_e d \Omega_e} \supset 
 { \kappa E_e |\boldsymbol{k}_e|  (E_0 - E_e)^2 \over    16 \pi^4  }  {\boldsymbol{J}  \boldsymbol{k}_e  \over J  E_e }  \xi A$ in the differential distribution obtained after integrating \cref{eq:CORR_distribution-restframe} over the neutrino angles. 
 Similarly, the neutrino asymmetry (capital $B$) is defined by 
 $ { d \Gamma \over  d E_e d \Omega_\nu   } \supset 
 { \kappa  E_e |\boldsymbol{k}_e|  (E_0 - E_e)^2 \over     16 \pi^4  } { \boldsymbol{J}  \boldsymbol{k}_\nu   \over J  E_\nu }  \xi B$ in  the differential distribution obtained after integrating \cref{eq:CORR_distribution-restframe} over the beta particle angles.
 In general, all these correlation coefficients are functions of the beta energy $E_e$, however, given the expression for the form factors in \cref{eq:DIST_formfactors}, at leading order in recoil they are constant. 
  For unpolarized decays, only the first line in \cref{eq:CORR_distribution-restframe} is relevant. 
 Note that, at recoil order, the full distribution in  \cref{eq:CORR_distribution-restframe} contains more information than that summarized in the correlation coefficients in \cref{eq:DIST_correlations}. 

Finally, integrating \cref{eq:CORR_distribution-restframe} over all angular variables one obtains the beta spectrum: 
\begin{align}
\label{eq:DIST_betaspectrum}
{ d \Gamma \over  d E_e   }=
 { \kappa  E_e |\boldsymbol{k}_e|  (E_0 - E_e)^2 \over 4 \pi^3 }
\bigg \{   \bigg ( 1   + {  3 E_e  \over   m_{\cal N} }    \bigg )  {\cal N } 
      - { E_e - {m_e^2 \over E_e} \over   m_{\cal N} }   \aleph    \bigg \} 
      + \cO(m_{\cal N}^{-2} )
. \end{align} 
This depends only on two form factors: 
${\cal N}$ and $\aleph$. 
At leading order, the curly bracket in \cref{eq:DIST_betaspectrum} contains only constant pieces, or, in the presence of exotic currents, terms proportional to ${m_e \over E_e}$ (the Fierz term). 
At first recoil order (that is $\cO(m_{\cal N}^{-1})$), also terms linear in $E_e$ appear.

\section{Discussion} 
\label{sec:DISC}

In this paper we have developed a new, convenient, fully  Lorentz-covariant formalism to deal with allowed beta decay amplitudes. 
We have followed the EFT approach, consistently expanding the amplitudes in powers of the recoil momentum $q$.  
As a simple application, we derived the fully differential decay rate (after summing over beta and daughter polarizations) to leading order 
($\cO(q^0)$) and including the complete first-order $(\cO(q^1)$) recoil corrections.  
Of course, not all the results regarding the differential distribution are new.
The leading-order result was derived by Jackson, Treiman, and Wyld in Ref.~\cite{Jackson:1957zz}. 
Their parameters $C_X$ and $C_X'$ are related to our Wilson coefficients by 
$C_{V,S}^i = C_{V,S} + C_{V,S}'$,
$C_{A,T}^i = \sqrt{J \over J+1} (C_{A,T} + C_{A,T}')$ (the combinations $C_{X} - C_{X}'$ have no counterparts in our paper because we do not consider right-handed neutrinos). 

Furthermore, effects due to complete recoil corrections to standard currents were calculated by Holstein in the seminal  Ref.~\cite{Holstein:1974zf}. 
For mixed Fermi-Gamow-Teller transitions, the results displayed in \cref{eq:DIST_formfactors} (in the limit where exotic currents are neglected) 
are equivalent up to $\cO(q^2)$ to the ones in Ref.~\cite{Holstein:1974zf} upon the following mapping of the parameters: 
 \begin{equation}  
 \label{eq:DISC_HolsteinMap} 
\begin{array}{c|c|c}
   & \beta^- & \beta^+  \\ \hline
 x_N a  &  C_V^i   &  \bar C_V^i  
 \\
x_N  c &  -  \sqrt{ J + 1 \over J} \bigg \{ 
 \bigg ( 1 +{E_0 \over 2 m_{\cal N} } \bigg )  C_A^i   
 + {E_0 \over   2 m_{\cal N} } C_{IT}^i \bigg \}   
 &  -  \sqrt{ J + 1 \over J} \bigg \{ 
 \bigg ( 1 +{E_0 \over 2 m_{\cal N} } \bigg )  \bar C_A^i  
 - {E_0 \over   2 m_{\cal N} } \bar C_{IT}^i \bigg \}  
 \\ 
x_N  b  &  \sqrt{ J + 1 \over J}  C_M^i   
&   \sqrt{ J + 1 \over J}  \bar C_M^i 
\\ 
x_N d &  -   \sqrt{ J + 1 \over J}  C_{IT}^i  
&  \sqrt{ J + 1 \over J}  \bar C_{IT}^i 
 \\ 
x_N  e &    C_{IS}^i  &  -  \bar C_{IS}^i 
\\ 
x_N f &    \sqrt{ (2J-1)(2J+3) \over 6 J^2}  \sqrt{ J + 1 \over J}  C_Q^i  &
-   \sqrt{ (2J-1)(2J+3) \over 6 J^2}  \sqrt{ J + 1 \over J}  \bar C_Q^i \\ 
\end{array}
,   \end{equation}  
where the common normalization factor is 
$x_N \equiv  {V_{ud} \over v^2 M_F^\mp}$.\footnote{
The agreement with Ref.~\cite{Holstein:1974zf} is up to a couple of apparent typos in that reference, some of which were noted previously. 
For example, the spin-dependent factor $\lambda_{J,J}$, equivalent to $\lambda_3$ in \cref{tab:CORR_GroupTheoreticalLambda}, should be divided by $-5\sqrt 6$~\cite{Melconian:RecoilOrder}. 
We also identified typos in the expressions for the spectral functions $F_6$, $F_8$, $F_{10}$, $F_{12}$, and $F_{14}$ in Ref.~\cite{Holstein:1974zf}, which are related to our form factors ${\cal B}$, ${\cal B}'$, ${\cal D}$, ${\cal C}_3'$, ${\cal C}$, and ${\cal C}_4'$}.   

It is also instructive to compare our formalism to that of Ref.~\cite{Falkowski:2021vdg}.    
That reference derives $\beta$ decay amplitudes from a nucleon-level  NR-EFT Lagrangian organized as an expansion in $\boldsymbol{\nabla}/m_N$, where $\boldsymbol{\nabla}$ is the spatial derivative, and $m_N \equiv {m_n + m_p \over 2}$ is the nucleon mass. 
This is qualitatively distinct from the nucleus-level EFT employed in this work, with a different set of strengths and limitations. 
The starting point in~\cite{Falkowski:2021vdg} is the  nucleon-level pionless NR-EFT Lagrangian with zero- and one-derivative interactions contributing to $\beta$ decay:\footnote{%
We omit here the Wilson coefficient $C_{E'}^+$ present in the Lagrangian of~\cite{Falkowski:2021vdg} because at this order it is redundant and can be absorbed into $C_A^+$.}
\begin{align}
\label{eq:CORR_LEFT-NR}  & 
 \cL_{\rm NR-EFT}   =   
 - (\psi_p^\dagger \psi_n)   \big  [    
C_V^+   (\bar e \bar \sigma^0 \nu)    +   C_S^+ (e^c \nu) 
  \big ] 
+  (\psi_p^\dagger \sigma^k \psi_n)   \big  [    
  C_A^+  (\bar e \bar \sigma^k \nu)  +   C_T^+  (e^c \sigma^0 \bar \sigma^k \nu)   \big ] 
  \nnl & 
   + {1 \over 2 m_N} \bigg \{  
i C_P^+   ( \psi_p^\dagger \sigma^k\psi_n) \nabla_k (e^c \nu)  
- C_M^+ \epsilon^{ijk}  ( \psi_p^\dagger \sigma^j \psi_n)   
\nabla_i  (\bar e \bar \sigma^k \nu)  
- i C_E^+  ( \psi_p^\dagger \sigma^k\psi_n) \nabla_k  (\bar e \bar \sigma^0 \nu)  
 \nnl  &
- i  C^+_{T1}  ( \psi_p^\dagger \psi_n) 
\nabla_k   (e^c \sigma^0 \bar \sigma^k \nu)  
+ i C^+_{T2} (\psi_p^\dagger \psi_n)  
(e^c  \overleftrightarrow{\partial_t} \nu) 
+ 2 i C^+_{T3} (\psi_p^\dagger \sigma^k \psi_n)   
(e^c  \overleftrightarrow{\nabla}_k \nu) 
\nnl & 
-  i C^+_{FV}  (\psi_p^\dagger  \overleftrightarrow{\nabla}_k \psi_n)   
(\bar e \bar \sigma^k \nu)  
+   i C^+_{FA}   (\psi_p^\dagger \sigma^k  \overleftrightarrow{\nabla}_k \psi_n)   
 (\bar e \bar \sigma^0 \nu)
+ C^+_{FT} \epsilon^{ijk} (\psi_p^\dagger \sigma^i  \overleftrightarrow{\nabla}_j \psi_n) 
 (e^c \sigma^0 \bar \sigma^k \nu) 
\bigg \}  
\nnl & 
+ \cO(\nabla^2)  + \hc 
, \end{align}
where $f_1 \Gamma \overleftrightarrow{\nabla} f_2    \equiv   f_1   \Gamma \nabla  f_2  - \nabla f_1 \Gamma f_2$, the nucleon degrees of freedom are described by non-relativistic quantum fields $\psi_N$, $N=p,n$, while leptons are described by the relativistic fields $\nu$ and $e$, $e^c$. 
Unlike $C_X^i$ in the covariant formalism, the Wilson coefficients $C_X^+$ in \cref{eq:CORR_LEFT-NR} do not carry the nuclear-transition-dependent label $i$, as they are specifically related to neutron interactions. 
They can be matched to the parameters of the quark-level EFT in \cref{eq:EFT_Lweft}, with the non-perturbative information about matrix elements of quark currents between nucleon states summarized in the charges $g_X$~~\cite{Falkowski:2021vdg}, many of which are known from lattice QCD or constrained by symmetries. 
In the NR-EFT approach the transition dependence enters via the matrix elements of nucleon bilinears between the nuclear states.   

We can match our transition-dependent Wilson coefficients $C_X^i$ to the nucleon-level parameters by equating beta decay amplitudes calculated in the two theories. 
The resulting map is given in \cref{eq:EFT_MaptoNR}.  
Then we can use the map in Ref.~\cite{Falkowski:2021vdg} between $C_X^+$ and the quark-level parameters to connect $C_X^i$ to the more fundamental parameters in \cref{eq:EFT_Lweft}. 
The result is 
  \begin{align} 
\label{eq:DISC_Map-CXiToEpsilon}
C_V^i  \simeq  & 
{V_{ud}   \over v^2} 
g_V \big ( 1 +  \epsilon_L + \epsilon_R \big ) \bigg ( 1 + {\Delta \over 2 m_{\cal N}  } \bigg )   
\sqrt{1 + \Delta_R^V}  
  , \nnl  
C_A^i   \simeq  & -  {V_{ud} \over v^2}      
  \sqrt{J \over J+1}  r_i
\bigg ( g_A -  { \Delta  \over  2 m_N } g_{IT}  \bigg )  
  \big ( 1 +  \epsilon_L - \epsilon_R \big ) \sqrt{1 + \Delta_R^A}  
 , \nnl 
  C_S^i \simeq  &
  {V_{ud} \over v^2}  g_S \epsilon_S  
, \nnl 
    C_T^i   \simeq   & 
   {V_{ud} \over v^2}    \sqrt{J \over J+1} r_i     g_T \epsilon_T   \bigg ( 1 +   {\Delta  \over  2   m_{\cal N} }   \bigg ) 
  , \nnl 
   C_M^i   \simeq  & 
    {V_{ud} \over v^2}    \sqrt{J \over J+1}  r_i   A
\big (  g_M +  \beta_i  g_V \big )  \big ( 1 +  \epsilon_L + \epsilon_R \big ) 
       , \nnl  
C_P^i   \simeq  &  
 {V_{ud} \over v^2}  \sqrt{J \over J+1}    r_i   A g_P \epsilon_P
     , \nnl 
  C_{T(1)}^i  \simeq  & 
  {V_{ud} \over v^2} A  \big (    g_T     -  2 g_T^{(1)}     -   \alpha_i  g_T   \big )   \epsilon_T  
, \nnl  
 C_{IS}^i  \simeq &
   {V_{ud} \over v^2}  A g_{IS}   \big ( 1 +  \epsilon_L + \epsilon_R \big ) 
, \nnl 
  C_{IT}^i   \simeq  & 
-    {V_{ud} \over v^2} \sqrt{J \over J+1}  r_i  A  g_{IT}   \big ( 1 +  \epsilon_L - \epsilon_R \big ) 
, \nnl   
C_{T(2)}^i  \simeq  & 
4      {V_{ud} \over v^2}   \sqrt{J \over J+1} r_i   A  g_T^{(3)}    \epsilon_T 
.  \end{align}   
Above, the transition-dependent parameters 
$r_i$, $\alpha_i$, and $\beta_i$ parametrize matrix elements of nucleon operators, which are defined in \cref{eq:EFT_NRmatrixelements}. 
$r_i$ is often called the {\em ratio of the Gamow–Teller to Fermi matrix elements}, while the other two appear in subleading matrix elements. 
The nucleon charges $g_X$ are defined by the matrix elements of quark bilinears between nucleon states, see \cref{eq:EFT_matrixelements-neutron}. 
For the subleading Wilson coefficients we only retain the leading recoil order matching, while for the leading ones we also track first recoil corrections. 
For $C_V^i$ and $C_A^i$ we display for completeness the radiative corrections to the matching parametrized by 
$\Delta_R^V$~\cite{Gorchtein:2023srs} and 
$\Delta_R^A$~\cite{Cirigliano:2022hob}, while such effects are ignored for the remaining Wilson coefficients.  
The $\simeq$ sign signals that these matching equations hold up to isospin-breaking corrections to the nuclear matrix elements of nucleon bilinears,  which are not included in Ref.~\cite{Falkowski:2021vdg}. 
They also do not take into account two-body current effects~\cite{Falkowski:2021vdg}. 
Nevertheless, they give an approximate picture of the expected transition dependence of the nuclear-level Wilson coefficients $C_X^i$. 
In particular, $C_V^i$ and $C_S^i$ are approximately the same for all Fermi or mixed Fermi-Gamow-Teller transitions, 
while the same parameter $r_i$ controls the leading order  axial and  tensor coefficients, so that their ratio is approximately fixed for all beta decays:  
$C_T^i/C_A^i \approx - {g_T \epsilon_T 
\over g_A ( 1 + \epsilon_L - \epsilon_R)}$. 
On the other hand, the ratio of pseudoscalar to axial coefficients in this approximation is proportional to the mass number $A$:  
$C_P^i/C_A^i \approx - A { g_P \epsilon_P 
\over g_A ( 1 + \epsilon_L - \epsilon_R)}$.  
Similar scaling is present for the coefficient responsible for weak magnetism~\cite{Severijns:2021lsp}: 
$C_M^i/C_A^i \approx - A {g_M + \beta_i g_V \over g_A}$ in the SM limit $\epsilon_X = 0$.
For mirror decays we can use the isospin symmetry prediction for $C_M^i$ together with phenomenological determinations  of $C_A^i$ to extract 
$\beta_i$, confirming it is an $\cO(1)$ parameter for a wide range of transitions, so that it  does not affect the overall scaling. 
The overall scale of the Wilson coefficients is set by 
${1 \over v^2}$ at leading order, 
and ${A \over v^2} = {m_{\cal N} \over m_N v^2}$ at subleading order, demonstrating that the dimensionless expansion parameter of the EFT is really the recoil momentum over the nucleon (rather than nuclear) mass $m_N$.

\section{Conclusions and outlook} 
\label{sec:CON}

In this paper we have proposed a novel treatment of nuclear beta decay.  
It makes use of the on-shell spinor-helicity formalism for massive and massless particles to conveniently handle arbitrary spins of the parent and daughter nuclei.
The formalism allows one to write down beta decay amplitudes in a manifestly Lorentz-invariant and little-group-covariant fashion.
From these, the usual observables such as the half-life, beta spectrum, and angular correlations can be readily obtained. 
By construction, the observables can be calculated in any reference frame, not only in the rest frame of the parent nucleus. 

Applied to allowed beta decay, we have employed an EFT approach, systematically expanding the amplitudes through linear order in the recoil momentum $q$ (where 
the dimensionless expansion parameters are $q/m_{\cal N}$ and  $q/m_N$,  with $m_{\cal N}$ and $m_N$  being the parent nucleus and nucleon mass). 
The most general form of the hadronic currents (which fully determine the amplitudes via \cref{eq:FORM_Mbeta-current}) to that order is given in \cref{eq:ABD_jXminus-expanded,eq:ABD_jXplus-expanded}. 
In this formulation,  the complete spin dependence of the currents relevant at this order is covariantly encoded in rank-1 and rank-2 tensors $S^\mu$  and $S^{\mu\nu}$, which are built from the massive spinors characterizing the parent nucleus. 
In the rest frame these reduce to more familiar spin matrices which can be built from Clebsch-Gordan coefficients following the usual rules of spin algebra. 
In that frame, the Wilson coefficients of the standard currents are in one-to-one correspondence with the $a \dots f$ parameters of Ref.~\cite{Holstein:1974zf}, cf.~\cref{eq:DISC_HolsteinMap}. 
The complete characterization of exotic (scalar, pseudoscalar, tensor)  currents at the first recoil order is new. 

Using our formalism, we calculated the fully differential distribution for allowed beta decay after summing over the polarizations of the daughter nucleus and the beta particle. 
That distribution can be characterized in a Lorentz-invariant way by a set of scalar form factors defined in \cref{eq:DIST_Msquared-parametrization}, which are functions of the kinematic invariants. 
In the rest frame one recovers the known spectral functions of  Ref.~\cite{Holstein:1974zf} in the limit where exotic currents are set to zero.
We also connect our covariant formalism to the NR-EFT approach of Ref.~\cite{Falkowski:2021vdg}. This in particular makes transparent the leading transition dependence of the Wilson coefficients, cf.~\cref{eq:DISC_Map-CXiToEpsilon}. 

There are several directions to go from here. 
It would be interesting to link our Wilson coefficients to nuclear-structure calculations, enabling systematic predictions for the coefficients. 
Furthermore, the on-shell spinor-helicity formalism is very convenient for higher-order calculations. 
Our approach can be readily  extended to the second order in recoil. 
Moreover, we can tap into the toolbox of (generalized) unitarity to calculate loop corrections to the differential distributions (see Ref.~\cite{Alviani:2024fby} 
for an application to the calculation of the Coulomb corrections to the $D$ parameter). 
Finally, working along similar lines,  the covariant formalism leads to a particularly transparent and convenient formulation of forbidden beta decays~\cite{inprogress}, where fewer results are currently known in the literature.

\section*{Acknowledgements}

We are grateful to Mart{\'\i}n Gonz{\'a}lez-Alonso  and  Ajdin Palavri{\'c} for useful discussions and comments on the manuscript. 
SCA is supported by the Generalitat Valenciana (Spain) through the plan GenT program (CIDEGENT/2021/037) and by  MCIU/AEI/10.13039/501100011033 (grants CEX2023-001292-S and PID2023-146220NBI00).
AF has received funding from the IN2P3 (CNRS) Master Project {\em On-shell techniques for forbidden nuclear decay}.

\appendix

\section{Useful spinor identities} 
\label{app:SI}

\subsection{Spin vector and tensors} 

In our conventions the spin pseudovector $S^\mu$ is a matrix in the little group space. 
For $J \geq 1/2$ it is defined  as 
 \begin{align} 
 \label{eq:SI_spinVector}
[S^\mu]_{O_1 \dots O_{2J} }^{I_1 \dots I_{2J} }  \equiv  &
 - {J \over 2 m_{\cal N} } 
 \big [  \chi_{1\, O_1}  \sigma^\mu    \tilde   \chi_1^{I_1}   +   \tilde  \chi_{1\, O_1}  \bar \sigma^\mu   \chi_1^{I_1} \big ]  
 \delta_{O_2}^{I_2} \dots  \delta_{O_{2J}}^{I_{2J}}
. \end{align} 
Here and everywhere below,  the upper and lower little group indices are, independently, implicitly fully symmetrized.
The spin vector is transverse: 
$p_1^\mu S_\mu = 0$, 
and satisfies 
$S^\mu S_\mu = - J(J+1) \mathbbm{1}$. 
This object encodes information about the polarization of the parent nucleus.

For $J \geq n/2$ we can also define a family of rank-n Lorentz tensors transforming covariantly under the little group. 
For our purpose we explicitly need the rank-two and rank-three tensors: 
\begin{align}
 \label{eq:SI_spinTensor}
\, [S^{\mu \nu}]_{O_1 \dots O_{2J} }^{I_1 \dots I_{2J} } \equiv &
     { J \big (J - {1\over 2} \big ) 
     \over (2 m_{\cal N})^2  }
   \big [  \chi_{1 \, O_1}  \sigma^\mu    \tilde   \chi_1^{I_1}   +   \tilde  \chi_{1\, O_1}  \bar \sigma^\mu   \chi_1^{I_1} \big ] 
  \big [  \chi_{1\, O_2}  \sigma^\nu    \tilde   \chi_1^{I_2}   +   \tilde  \chi_{1\, O_2}  \bar \sigma^\nu   \chi_1^{I_2} \big ] 
\delta_{O_3}^{I_3} \dots  \delta_{O_{2J}}^{I_{2J}} 
\nnl &  
  +  {J \big (J - {1\over 2} \big ) \over 3}  
\Pi^{\mu \nu}    \delta_{O_1}^{I_1} \dots  \delta_{O_{2J}}^{I_{2J}}, 
  \nnl 
\, [S^{\mu \nu \rho}]_{O_1 \dots O_{2J} }^{I_1 \dots I_{2J} }  \equiv & 
- {J (J-1/2) (J-1)  \over (2 m_{\cal N})^3  }
     \big [  \chi_{1 \, O_1}  \sigma^\mu    \tilde   \chi_1^{I_1}   +   \tilde  \chi_{1\, O_1}  \bar \sigma^\mu   \chi_1^{I_1} \big ] 
  \big [  \chi_{1\, O_2}  \sigma^\nu    \tilde   \chi_1^{I_2}   +   \tilde  \chi_{1\, O_2}  \bar \sigma^\nu   \chi_1^{I_2} \big ]  
  \nnl \times &
    \big [  \chi_{1\, O_3}  \sigma^\rho   \tilde   \chi_1^{I_3}   +   \tilde  \chi_{1\, O_3}  \bar \sigma^\rho   \chi_1^{I_3} \big ]  
\delta_{O_4}^{I_4} \dots  \delta_{O_{2J}}^{I_{2J}} 
 \nnl &
   + {(J-1)(J-1/2) \over 5}  \bigg [ \Pi^{\mu \nu}   S^\rho   +  \Pi^{\rho \mu}   S^\nu +  \Pi^{\nu\rho}   S^\mu \bigg ]_{O_1 \dots O_{2J} }^{I_1 \dots I_{2J} }
, \end{align} 
where $ \Pi^{\mu \nu}  \equiv \eta^{\mu \nu} - {p_1^\mu p_1^\nu \over m_{\cal N}^2 }$. 
These are symmetric in all Lorentz indices. 
Much as the spin vector, the spin tensors are  transverse: 
$p_1^{\mu_i} S_{\mu_1 \dots \mu_i \dots  \mu_n} =  0$. 
Furthermore, they are traceless: 
$\eta^{\mu_i \mu_j} S_{\mu_1 \dots \mu_i \dots \mu_j \dots \mu_n} = 0$. 

For the sake of calculating differential distributions, we use the following rules for multiplying spin vectors and tensors:
 \begin{align}   
S^\mu  S^\nu  =  & 
 -  {J (J+1) \over 3}  \Pi^{\mu \nu}    \mathbbm{1}  
 +   {i     \over 2 m_{\cal N} }   
 \epsilon^{\mu \nu \alpha \beta}  p_1{}_\alpha S_\beta 
+ S^{\mu \nu}
\nnl 
 S^\mu  S^{\nu \rho} =  & 
  -    { (2J-1)(2J+3) \over 20  }   \bigg ( 
 \Pi^{\mu\nu} S^\rho  +     \Pi^{\mu\rho}   S^\nu  
- { 2 \over 3}    \Pi^{\nu\rho}  S^\mu   \bigg )   
\nnl  &  
-  {i \over 2 m_{\cal  N} } \epsilon^{\mu \nu \alpha \beta}  p_1{}_\alpha  S_\beta{}^\rho 
-  {i \over 2 m_{\cal  N} } \epsilon^{\mu \rho \alpha \beta}  p_1{}_\alpha  S_\beta{}^\nu
+  S^{\mu\nu \rho} 
,    \end{align} 
where $[S^{\mu_1 \dots}  S^{\nu_1 \dots } ]_{O_1 \dots O_{2J} }^{I_1 \dots I_{2J} }\equiv
\sum_{K_1 \dots K_{2J}=1}^2 
[S^{\mu_1 \dots}]_{K_1 \dots K_{2J} }^{I_1 \dots I_{2J} }
[S^{\nu_1 \dots } ]_{O_1 \dots O_{2J} }^{K_1 \dots K_{2J} }$. 

\subsection{Leptonic sum rules} 
Here we present the leptonic sum rules used for calculating differential distributions. 
Summation $\sum_{K=1}^2$ of the  little group index $K$  is implicit on the left-hand side of all equations,  which corresponds to summing over two polarizations of the beta particle. 

  \begin{align}
\label{eq:LSR_sumrule-VV}
   (\tilde \chi_{3}{}_K \bar   \sigma^\mu  \lambda_4  )   (\tilde \chi_{3}{}_K  \bar   \sigma^\nu  \lambda_4  )^*  = & 
2 \big [ p_3^\mu p_4^\nu + p_3^\nu p_4^\mu - \eta^{\mu \nu} (p_3 p_4) 
- i \epsilon^{ \mu \nu \alpha\beta }   p_3{}_\alpha p_4{}_\beta  \big ] 
, \nnl 
   (\tilde \chi_{3}^K \bar   \sigma^\mu  \lambda_4  )^*   (\tilde \chi_{3}^K  \bar   \sigma^\nu  \lambda_4  ) = &  
2 \big [ p_3^\mu p_4^\nu + p_3^\nu p_4^\mu - \eta^{\mu \nu} (p_3 p_4) 
+  i \epsilon^{ \mu \nu \alpha\beta }   
p_3{}_\alpha p_4{}_\beta  \big ] 
   . \end{align} 
\begin{align}
  \label{eq:LSR_sumrule-VS} 
   (\tilde \chi_{3}{}_K \bar   \sigma^\mu  \lambda_4  )   (  \chi_{3}{}_K \lambda_4 )^*    = & 
  -2 m_e p_4^\mu 
 , \nnl 
 (\tilde \chi_{3}^K \bar   \sigma^\mu  \lambda_4  )^*   (  \chi_{3}{}^K \lambda_4 )   = & 
   -2 m_e p_4^\mu 
   . \end{align}  
\begin{align}
\label{eq:LSR_sumrule-VT}   
 (\tilde \chi_{3}{}_K \bar   \sigma^\mu \lambda_4  )  
 (\chi_{3}{}_K \sigma^{\alpha\beta} \lambda_4 )^*      = & 
  2 i m_e   \big [
\eta^{\mu\alpha } \eta^{\beta \rho} - \eta^{\mu\beta } \eta^{\alpha \rho}  + i \epsilon^{\mu\alpha \beta \rho} 
\big  ] p_4{}_\rho
, \nnl 
 (\tilde \chi_{3}^K \bar   \sigma^\mu \lambda_4  )^*  (\chi_{3}{}^K \sigma^{\alpha\beta} \lambda_4 )
 = & 
   2 i m_e   \big [
- \eta^{\mu\alpha } \eta^{\beta \rho} + \eta^{\mu\beta } \eta^{\alpha \rho}  + i \epsilon^{\mu\alpha \beta \rho} 
\big  ] p_4{}_\rho
 ,   \end{align}  

\begin{align} 
\label{eq:LSR_sumrule-SS} 
 (  \chi_{3}{}_K \lambda_4 )    (  \chi_{3}{}_K \lambda_4 )^*     = &  
   (2 p_3 p_4)
   , \nnl 
    (  \chi_{3}^K \lambda_4 )^*    (  \chi_{3}^K \lambda_4 )    = &  
   (2 p_3 p_4)
    . \end{align}  
\begin{align} 
\label{eq:LSR_sumrule-TS}  
 (\chi_{3K}  \sigma^{\mu \nu} \lambda_4 )  (  \chi_{3K} \lambda_4 )^* = & 
2 i  \big [ p_3^\mu p_4^\nu  - p_3^\nu p_4^\mu -  i \epsilon^{\mu\nu\alpha\beta}   p_3{}_\alpha p_4{}_\beta   \big ] 
, \nnl 
 (\chi_{3}^K   \sigma^{\mu \nu} \lambda_4 )^* (  \chi_{3}^K \lambda_4 ) = & 
 2 i  \big [ - p_3^\mu p_4^\nu  + p_3^\nu p_4^\mu -  i \epsilon^{\mu\nu\alpha\beta}   p_3{}_\alpha p_4{}_\beta   \big ] 
   . \end{align}  
\begin{align}  
\label{eq:LSR_sumrule-TT}   
(\chi_3{}_K \sigma^{\mu\nu} \lambda_4) ( \chi_3{}_K  \sigma^{\alpha\beta}  \lambda_4)^*  =  & 
 2  \bigg \{  
 \eta^{\mu\beta}       p_3^\alpha p_4^\nu 
  +   \eta^{\nu\alpha}       p_3^\beta p_4^\mu 
 - \eta^{\nu\beta}    p_3^\alpha p_4^\mu 
  -  \eta^{\mu\alpha}     p_3^\beta p_4^\nu 
  \nnl    &
 +  \eta^{\mu \beta}    p_3^\nu p_4^\alpha  
 + \eta^{\nu  \alpha}  p_3^\mu p_4^\beta 
   - \eta^{\nu   \beta } p_3^\mu p_4^\alpha 
 - \eta^{\mu \alpha}   p_3^\nu p_4^\beta 
\nnl    &
+   (p_3 p_4) \big [ 
 \eta^{\mu \alpha}  \eta^{\nu  \beta}  - \eta^{\mu \beta}  \eta^{\nu  \alpha}  \big ] 
 \nnl    &
 +  i \epsilon^{\mu\nu A \beta} p_4{}_A p_3^\alpha   
- i \epsilon^{\mu\nu A \alpha} p_4{}_A p_3^\beta    
+ i \epsilon^{ \nu \alpha  \beta  A }  p_3{}_A   p_4^\mu 
- i \epsilon^{ \mu \alpha  \beta  A } p_3{}_A  p_4^\nu   
\bigg \} 
, \nnl 
(\chi_3{}^K \sigma^{\mu\nu} \lambda_4)^*  ( \chi_3^K  \sigma^{\alpha\beta}  \lambda_4) =  & 
- 2  \bigg \{  
 \eta^{\nu\beta}    p_3^\alpha p_4^\mu 
  -  \eta^{\mu\beta}       p_3^\alpha p_4^\nu 
  -   \eta^{\nu\alpha}       p_3^\beta p_4^\mu 
  +  \eta^{\mu\alpha}     p_3^\beta p_4^\nu 
  \nnl    &
   +\eta^{\nu   \beta } p_3^\mu p_4^\alpha 
    - \eta^{\mu \beta}    p_3^\nu p_4^\alpha  
-   \eta^{\nu  \alpha}  p_3^\mu p_4^\beta 
 +\eta^{\mu \alpha}   p_3^\nu p_4^\beta 
\nnl    &
+  (p_3 p_4) \big [ \eta^{\mu \beta}  \eta^{\nu  \alpha}
 - \eta^{\mu \alpha}  \eta^{\nu  \beta} \big ] 
 \nnl    &
- i \epsilon^{\mu\nu A \alpha} p_4{}_A p_3^\beta    
+  i \epsilon^{\mu\nu A \beta} p_4{}_A p_3^\alpha   
- i \epsilon^{ \mu \alpha  \beta  A } p_3{}_A  p_4^\nu   
+ i \epsilon^{ \nu \alpha  \beta  A }  p_3{}_A   p_4^\mu 
\bigg \} 
. \end{align}

\subsection{Recoil expansion of spinor bilinears} 

For $q^2  < 4m_{\cal N}^2$, the massive spinors describing the two nuclei can be related as~\cite{Aoude:2020onz,Cangemi:2023ysz}  
\begin{align}  
\label{eq:SI_chi2ofchi1}
\chi_2 = &  - {1 \over \sqrt {1 -  {q^2 \over 4m_{\cal N}^2}  }} \bigg  [  \chi_1  -  {q \sigma \over 2 m_{\cal N} } \tilde \chi_1 \bigg ] 
 , \nnl 
\tilde \chi_2 = & {1 \over \sqrt {1 -  {q^2 \over 4m_{\cal N}^2} }}  \bigg [   \tilde  \chi_1 -   {q \bar \sigma \over 2 m_{\cal N} }  \chi_1  \bigg ] 
.  \end{align}
This relation implies identification of the little group frames of the two particles. 
It allows one to expand spinor bilinears  in powers of $q$: 
\begin{align}   
\label{eq:SI_RecoilExpansion}
\chi_2 \chi_1 +  \tilde \chi_2 \tilde \chi_1  = & 
2 m_{\cal N}  \boldsymbol{1}    + \cO(q^2)
,\nnl 
\chi_2 \chi_1 - \tilde \chi_2 \tilde \chi_1   = &  
-{1\over 2 m_{\cal N}}  \big [  \chi_1 q\sigma    \tilde \chi_1  
+  \tilde \chi_1  q \bar \sigma  \chi_1  \big ]  + \cO(q^2) 
,\nnl
\chi_2 \sigma_\mu  \tilde  \chi_1 +  \tilde \chi_2 \bar \sigma_\mu \chi_1  =  & 
  (p_1 - p_2)_\mu  \boldsymbol{1}   
 - {i  \over 2 m_{\cal N}^2 }   
\epsilon_{\mu \nu \alpha \beta} q^\nu p_1^\alpha  \big [ \chi_1\sigma^\beta  \tilde  \chi_1 
+   \tilde \chi_1 \bar \sigma^\beta   \chi_1 \big ]
+ \cO(q^2) 
, \nnl 
\chi_2 \sigma^\mu  \tilde  \chi_1 -  \tilde \chi_2 \bar \sigma^\mu \chi_1  = &  
- \big [ \chi_1  \sigma^\mu    \tilde   \chi_1    +   \tilde  \chi_1  \bar \sigma^\mu   \chi_1  \big ]  
-    {p_1^\mu  \over  2 m_{\cal N}^2} \big [\chi_1  q  \sigma     \tilde  \chi_1  +   \tilde \chi_1  q \bar \sigma   \chi_1  \big ] 
+ \cO(q^2) 
,\nnl 
\chi_2  \sigma^{\mu \nu} \chi_1  +    \tilde \chi_2 \bar \sigma^{\mu \nu} \tilde   \chi_1 
=  & 
   { 1 \over m_{\cal N} } \epsilon^{\mu \nu\alpha \beta } p_1{}_\alpha  
    \big [   \chi_1  \sigma_\beta      \tilde \chi_1  +  \tilde \chi_1  \bar \sigma_\beta   \chi_1   \big ] 
+ i {p_1^\mu  q^\nu -  p_1^\nu  q^\mu   \over m_{\cal N} } \boldsymbol{1}
  \nnl   - &
  {1 \over 2m_{\cal N} }\epsilon^{ \mu \nu \alpha \beta}  q_\alpha  \big [ 
\chi_1  \sigma_\beta  \tilde   \chi_1     +  \tilde \chi_1   \bar \sigma_\beta      \chi_1   \big ]  
+ \cO(q^2)  
 ,\nnl
\chi_2  \sigma^{\mu \nu} \chi_1   -    \tilde \chi_2 \bar \sigma^{\mu \nu} \tilde   \chi_1    =  & 
   { i \over m_{\cal N}}  \bigg \{ 
   p_1^\mu    \big [   \chi_1  \sigma^\nu      \tilde \chi_1  +  \tilde \chi_1  \bar \sigma^\nu   \chi_1    \big ] 
  -  p_1^\nu  \big [  \chi_1  \sigma^\mu   \tilde \chi_1  +   \tilde \chi_1 \bar  \sigma^\mu \chi_1   \big ]   
       \bigg\}   
      + {1 \over m_{\cal N} }\epsilon^{ \mu \nu \alpha \beta} p_1^\alpha  q^\beta   \boldsymbol{1} 
        \nnl  - &     {i \over 2 m_{\cal N} } \bigg \{ 
q^\mu \big [ \chi_1   \sigma^\nu   \tilde   \chi_1     +  \tilde \chi_1   \bar \sigma^\nu  \chi_1    \big ] 
- q^\nu \big [ \chi_1   \sigma^\mu   \tilde   \chi_1     +  \tilde \chi_1   \bar \sigma^\mu  \chi_1    \big ] 
\bigg \}  + \cO(q^2)  
,\nnl
\big[  \chi_2  \sigma^{\mu \nu}  \chi_1  +  \tilde \chi_2  \bar \sigma^{\mu \nu} \tilde  \chi_1  \big ]q^\nu  = & 
  { 1 \over m_{\cal N}  }  
\epsilon^{\mu \nu \alpha \beta} q_\nu p_1{}_\alpha       \big [ \chi_1\sigma_\beta  \tilde  \chi_1 
+   \tilde \chi_1 \bar \sigma_\beta   \chi_1 \big ] 
  + \cO(q^2) 
  ,\nnl
\big[  \chi_2  \sigma^{\mu \nu}  \chi_1 -  \tilde \chi_2  \bar \sigma^{\mu \nu} \tilde  \chi_1  \big ]q_\nu  = &  
{  i \over m_{\cal N}  }  p_1^\mu \big [ \chi_1 q \sigma \tilde  \chi_1 + \tilde  \chi_1q  \bar \sigma \chi_1  \big ]   + \cO(q^2) 
 ,  \end{align}  
where $\boldsymbol{1}   \equiv  \delta^I_O$, and spinors appearing to the left (right) in spinor contractions implicitly carry lower $O$ (upper $I$) little group indices.

\section{More fundamental EFTs}  
\label{app:EFT}

In this work we employ a general Lorentz-covariant parametrization of nuclear beta amplitudes in a systematic EFT expansion in the recoil momentum $q$. 
The Wilson coefficients $C_X^i$ of this nuclear-level EFT can be connected to more fundamental parameters of quark- and nucleon-level EFTs. 
Below we briefly summarize these more fundamental EFTs. 

\subsection{Quark-level EFT}

At scales above $\mu \simeq $ 2~GeV, quarks and gluons become perturbative degrees of freedom. 
The effective theory at that scale relevant for $\beta$ transitions, the so-called weak EFT (WEFT),   describes charged-current effective interactions between quark and lepton currents, organized in an expansion in $\partial/m_W$, where $m_W$ is the $W$ boson mass. 
The leading operators can be parametrized as~\cite{Bhattacharya:2011qm}
\begin{align}
\cL_{\rm WEFT}   \supset   & 
- \frac{V_{ud}}{v^2}  \Big \{
2 \left( 1 +  \epsilon_L \right) (\bar{e} \bar \sigma_\mu \nu) ( \bar u \bar \sigma^\mu d) 
+ 2 \epsilon_R   (\bar{e} \bar \sigma_\mu \nu)
 (u^c \sigma^\mu \bar d^c )
\nnl & 
+{1 \over 2} \epsilon_T  (e^c  \sigma_{\mu \nu} \nu)     
\big [ \bar u \bar \sigma^{\mu \nu} \bar d^c   + u^c  \sigma^{\mu \nu} d   \big ] 
+ \epsilon_S  (e^c \nu)  \big [ \bar u \bar d^c   + u^c  d    \big ]  
- \epsilon_P  (e^c \nu)   
 \big [ \bar u \bar d^c   -  u^c  d    \big ]  
\Big \} 
\nnl & 
+ {\rm h.c.}~ \qquad 
\label{eq:EFT_Lweft}
, \end{align}
where  $e$ and $\bar e^c$ describe left- and right-handed electrons, $u$, $\bar u^c$, $d$, $\bar d^c$ describe left- and right-handed up and down quarks, 
and $\nu$ describes the left-handed electron neutrino.
All these are fermionic quantum fields in the 2-component spinor notation~\cite{Dreiner:2008tw}. 
Furthermore, $V_{ud}$ is the CKM matrix element (with conventions fixed such that it is real), 
and the dimensionful scale is 
$v =  (\sqrt 2 G_F)^{-1/2} \approx 246.22$~GeV. 
The Wilson coefficients $\epsilon_X$, 
$X = L,R,S,P,T$, parametrize possible effects of non-SM particles heavier than 2~GeV.
In the SM limit,  $\epsilon_X = 0$ for all $X$. 
Given this Lagrangian, we can identify the quark-level hadronic currents appearing in \cref{eq:FORM_jX-definition}:  
\begin{align} 
\label{eq:EFT_jX}
\hat j_{V-}^\mu(x) = &  {V_{ud} (1+ \epsilon_L + \epsilon_R  )  \over v^2} \big [ 
u^c \sigma^\mu \bar d^c   +  \bar u \bar \sigma^\mu d 
  \big ]
, \quad 
\hat j_{V+}^\mu(x) =   {V_{ud} (1+ \bar \epsilon_L+ \bar \epsilon_R )  \over v^2} 
\big [ d^c \sigma^\mu \bar u^c   +  \bar d \bar \sigma^\mu u \big ] 
, \nnl
\hat j_{A-}^\mu(x) = &  {V_{ud} (1+ \epsilon_L - \epsilon_R )  \over v^2} 
\big [  u^c \sigma^\mu \bar d^c   -  \bar u \bar \sigma^\mu d    \big ]  
, \quad  
\hat j_{A+}^\mu(x) =   {V_{ud} (1+ \bar \epsilon_L - \bar \epsilon_R )  \over v^2} 
\big [ d^c \sigma^\mu \bar u^c   -  \bar d \bar \sigma^\mu u   \big ]  
, \nnl
\hat j_{S-}(x) = &  {V_{ud} \epsilon_S \over v^2} 
\big [  \bar u \bar d^c   +u^c  d    \big ] 
, \qquad  
\hat j_{S+}(x) =   {V_{ud} \bar \epsilon_S \over v^2} \big [ 
\bar d \bar u^c   + d^c  u \big ] 
, \nnl
\hat j_{P-}(x) = &   - {V_{ud} \epsilon_P \over v^2}  
\big [ \bar u \bar d^c   -  u^c  d  \big ] 
, \qquad  
\hat j_{P+}(x) =    {V_{ud} \bar \epsilon_P \over v^2} 
\big [  
\bar d \bar u^c   -  d^c  u \big ]  
, \nnl 
\hat j_{T-}^{\mu \nu}(x) = &  {V_{ud} \epsilon_T \over v^2} \big [
\bar u \bar \sigma^{\mu \nu} \bar d^c   + u^c  \sigma^{\mu \nu} d  \big ] 
, \qquad  
\hat j_{T+}^{\mu \nu}(x) =   {V_{ud} \bar \epsilon_T \over v^2} 
\big [ \bar d \bar \sigma^{\mu \nu} \bar u^c   + d^c  \sigma^{\mu \nu} u   \big ] 
.   \end{align}  
To calculate the neutron beta decay amplitude from the Lagrangian in \cref{eq:EFT_Lweft}, we need matrix elements of the quark currents between nucleon (neutron and proton) states. 
Their most general form allowed by the Lorentz symmetry and the discrete $P$ symmetry of QCD can be parametrized as~\cite{Weinberg:1958ut,Holstein:1974zf,Bhattacharya:2011qm}
\begin{align}
\label{eq:EFT_matrixelements-neutron}
\bra{p}  \bar{u}   \gamma_\mu  d  \ket{n}  =& 
  \bar  u_p \bigg [ g_V \gamma_\mu + { g_{IS}  \over 2 m_N } q_\mu 
- i {g_M - g_V \over 2 m_N} \sigma_{\mu \nu} q^\nu \bigg ] u_n 
+ \mathcal{O}(q^2) 
, \nnl 
\bra{p}  \bar{u}   \gamma_\mu  \gamma_5 d  \ket{n}  =& 
 \bar  u_p \bigg [ g_A \gamma_\mu \gamma_5  + { g_{IP}  \over 2 m_N }\gamma_5 q_\mu 
- i {g_{IT} \over 2 m_N} \sigma_{\mu \nu} \gamma_5 q^\nu \bigg ] u_n 
+ \mathcal{O}(q^2) 
, \nnl 
\bra{p}  \bar{u}   d  \ket{n}  =& g_S \bar u_p u_n  
+ \mathcal{O}(q^2)  
, \nnl 
\bra{p} \bar u \gamma_5  d  \ket{n}  =& 
g_P \bar  u_p \gamma_5 u_n
+ \mathcal{O}(q^2)  
,\nnl 
\bra{p}  \bar{u}   \sigma_{\mu\nu}  d  \ket{n}  =& 
\bar  u_p \bigg [ g_T \sigma_{\mu\nu}  
+  {i g_T^{(1)} \over m_N}  \big ( q_\mu \gamma_\nu -   q_\nu \gamma_\mu \big )  
\nnl  & 
+   {i g_T^{(3)} \over m_N}  \big ( \gamma_\mu \gamma_\rho  \gamma_\nu -  \gamma_\nu  \gamma_\rho  \gamma_\mu  \big )   q^\rho
 \bigg ] u_n 
 + \mathcal{O}(q^2)  
 . \end{align}
Above,  $u_N$ is the Dirac spinor wave function of the neutron or proton. 
Different Lorentz structures are weighted by nucleon charges $g_X$, most of which are known with sufficient precision.  
Isospin symmetry fixes $g_V = 1$ (up to corrections quadratic in isospin breaking~\cite{Ademollo:1964sr,Donoghue:1990ti}), and relates $g_M$ to the difference of the proton and neutron magnetic moments~\cite{GellMann:1964tf,Holstein:1974zf,Cirigliano:2022hob},  
${ g_M \over g_V } =\frac{\mu_{p}-\mu_{n}}{\mu_{N}}  \simeq 4.706$, 
where $\mu_{N}=\frac{e}{2 m_p}$ is the nuclear magneton. 
It also implies vanishing of the induced scalar  $g_{IS}$, induced tensor $g_{IT}$, and the subleading tensor $g_T^{(3)}$ charges, which are then expected to be $\cO(10^{-2})$ due to isospin breaking effects, see in particular Refs.~\cite{Shiomi:1996np,Sharma:2009hg}.
The axial, scalar, and tensor charges can be calculated on the lattice, with the Flavor Lattice Averaging Group (FLAG) 2024 average 
$g_A = 1.263(10)$, $g_S=1.085(114)$, 
$g_T = 0.981(21)$~\cite{FlavourLatticeAveragingGroupFLAG:2024oxs}.  
Using partial conservation of the axial current one finds 
$g_{P}=\frac{m_{N}}{\overline{m}_{q}}g_{A}= 346(9)$,  and 
$g_{IP}=-\frac{4 m_{N}^{2}}{m_{\pi}^{2}}g_{A} \approx -226(5)$~\cite{Gonzalez-Alonso:2013ura}, 
where $\overline{m}_{q}$ is the average of the light quark masses, and $m_\pi$ is the pion mass. 
We are not aware of a precise determination for $g_{T}^{(1)}$, but by naturalness considerations it is expected to be $\mathcal{O}(1)$.

\subsection{Nucleon-level EFT}

The nucleon-level NR-EFT approach in Ref.~\cite{Falkowski:2021vdg} is based on the effective Lagrangian in \cref{eq:CORR_LEFT-NR}. 
To calculate beta decay amplitudes from it,  one needs matrix elements of nucleon bilinears between nuclear states. 
For $\beta^-$ transitions, the parametrization given in  Ref.~\cite{Falkowski:2021vdg} in the rest frame of the parent nucleus and in the isospin-symmetric limit  reduces to\footnote{%
Compared to Ref.~\cite{Falkowski:2021vdg}, we replace $[{\cal T}_{(J)}^k]_{J_z'}^{\, J_z}$ with 
$\boldsymbol{S}^k$ and  $\boldsymbol{P}^k$ with  
$- \boldsymbol{q}^k$, as those equalities hold in the rest frame. 
We also rename $\alpha_{FT} \to \alpha_i$, and 
$\beta_{FV} \to \beta_i$, stressing transition-dependence of these parameters.  
Finally, we corrected the sign error in the first term in the curly bracket of the last matrix element in \cref{eq:EFT_NRmatrixelements}. 
}
\begin{align}
\label{eq:EFT_NRmatrixelements} 
\langle {\cal N}_2 | \psi_p^\dagger  \psi_n |{\cal N}_1  \rangle  = & 
2 m_{\cal N}  M_F^- \mathbbm{1}
 , \nnl 
\langle {\cal N}_2 | \psi_p^\dagger \sigma^k \psi_n  |{\cal N}_1  \rangle   = &    
2 m_{\cal N}  M_F^-  { r_i \over \sqrt{J(J+1)} }\boldsymbol{S}^k  
, \nnl 
\langle {\cal N}_2 | \psi_p^\dagger  \overleftrightarrow{\nabla}^k   \psi_n  |{\cal N}_1 \rangle   = &
2 m_{\cal N}   M_F^-
\bigg \{ - i  {1 \over A}  \boldsymbol{q}^k \mathbbm{1}
-  \beta_i {r_i \over \sqrt{J(J+1)} }   \epsilon^{klm}  \boldsymbol{q}^l \boldsymbol{S}^m 
 \bigg \} 
  , \nnl 
\langle {\cal N}_2 | \psi_p^\dagger \sigma^l \overleftrightarrow{\nabla}^l \psi_n  |{\cal N}_1  \rangle   = &   
2 m_{\cal N}   M_F^- \bigg \{ 
  -  i { r_i  \over A \sqrt{J(J+1)} }   \boldsymbol{q}^l \boldsymbol{S}^l
\bigg \} 
,\nnl 
\epsilon^{klm} \langle {\cal N}_2 | \psi_p^\dagger \sigma^l  \overleftrightarrow{\nabla}^m \psi_n |{\cal N}_1  \rangle   = & 
2 m_{\cal N}  M_F^- 
\bigg \{ 
 i  {r_i  \over A \sqrt{J(J+1)}}  \epsilon^{klm}   \boldsymbol{q}^l \boldsymbol{S}^m 
+    \alpha_i  \boldsymbol{q}^k  \mathbbm{1}
 \bigg \} 
. \end{align}
Given these matrix elements and the Lagrangian in \cref{eq:CORR_LEFT-NR}, we can calculate $\beta^-$ decay amplitudes in the rest frame of the parent nucleus. 
These can be compared with the amplitudes in \cref{eq:FORM_Mbeta-current} in the covariant formalism evaluated in the rest frame, with the currents parametrized in \cref{eq:ABD_jXminus-expanded}. 
The two are equal upon the following identification of the Wilson coefficients: 
  \begin{align}   
\label{eq:EFT_MaptoNR} 
C_V^i  = & 
  C_V^+   +    {\Delta \over 2 m_{\cal N}  }  C^+_{FV} - {m_e \over 2 m_N  }   C^+_{T2} 
  , \nnl  
  C_A^i   =  &  \sqrt{J \over J+1}  r_i
  \bigg (  C_A^+    
  -  {  m_e \over  m_N }    C^+_{T3}    \bigg ) 
 , \nnl 
   C_M^i     = & 
  \sqrt{J \over J+1}  r_i  A  
  \big ( C^+_{M} + \beta_i C^+_{FV} \big ) 
       , \nnl  
 C_{IT}^i    = & 
\sqrt{J \over J+1}  r_i   \big (C^+_{FA} - C_A^+ - A C_E^+ \big )  
  , \nnl 
 C_S^i  +   {m_e \over 2 m_{\cal N}}  C_{IS}^i  = &
 C_S^+  +  {m_e \over 2 m_{\cal N}}  \big ( C_V^+ -   C^+_{FV}  \big ) 
  , \nnl 
C_P^i    = &  
 \sqrt{J \over J+1}  r_i  A  C_P^+
   , \nnl 
 C_T^i    =  & 
  \sqrt{J \over J+1} r_i  \bigg (     C_T^+   
+   {\Delta  \over  2   m_{\cal N} }  C^+_{FT}   \bigg ) 
     , \nnl 
  C_{T(1)}^i  =  & 
 A  \big (  C^+_{T1}   +  C^+_{T2}  -   \alpha_i  C^+_{FT} \big )   
      , \nnl 
 C_{T(2)}^i  = & 
\sqrt{J \over J+1} r_i
\big (C^+_{FT} - C_T^+   -  2 A C^+_{T3}  \big )      
,   \end{align}  
while $C_Q^i = C_{TQ}^i = 0$ in the isospin-symmetric parametrization above. 
These last two Wilson coefficients are generated only when the matrix elements in \cref{eq:EFT_NRmatrixelements} are generalized to include isospin-breaking corrections. 
In particular, if we include 
$\langle {\cal N}_2| \psi_p^\dagger  \overleftrightarrow{\nabla}^k   \psi_n |{\cal N}_1 \rangle  \supset 
2 i m_{\cal N}  M_F^-  {\delta_i  \over J^2} \boldsymbol{q}^l  \boldsymbol{S}^{kl}$, 
and 
$\epsilon^{klm} \langle {\cal N}_2| \psi_p^\dagger \sigma^l  \overleftrightarrow{\nabla}^m \psi_n |{\cal N}_1 \rangle  = 
2 m_{\cal N}  M_F^- 
 {  \gamma_i \over J^2 }   \boldsymbol{q}^l \boldsymbol{S}^{kl}$ 
 then the quadrupole Wilson coefficients are generated as 
  \begin{align}   
  C_Q^i =   & 
    - A \delta_i    C^+_{FV} 
, \qquad 
C_{TQ}^i     =   
 A    \gamma_i C^+_{FT}  
 . \end{align}   

\section{Exotic recoil-order contributions} 
\label{app:ERC}

For allowed decays, the first-recoil-order  contributions  ($\cO(q)$) to the form factors from interference between standard and exotic currents take the form 
\begin{align}   
\label{eq:ERC_delta1}
\delta_1{\cal N } = &  
{m_e \over  m_{\cal N} }  \re \bigg \{ 
 \bigg ( {E_0 \over E_e }  - 1 \bigg ) 
 C_V^i  \bar C_{T(1) }^i
 \mp  C_S^i (\bar C_V^i \mp \bar C_{IS}^i)
 + {\lambda_0 \over 3}  \bigg [ 
 \bigg ( {E_0 \over E_e } - 1 \bigg )  C_A^i  \bar C_P^i
  \nnl &  
-  \bigg ( {E_0 \over E_e }  + 2  \bigg )  
C_A^i( \bar C_{T(2) }^i  \pm  \alpha_T \bar C_T^i )  
+ 2   \bigg ( {E_0 \over E_e }  - 1 \bigg )    C_T^i \bar C_M^i 
+    \bigg ( {E_0 \over E_e }  - 1 \bigg )    
C_T^i  ( \bar C_{IT}^i \pm \alpha_A \bar C_A^i )  
\bigg ]  \bigg \} 
, \nnl 
\delta_1\aleph = &
 {m_e \over  m_{\cal N} } \re \bigg \{  
 C_V^i \bar C_{T(1) }^i
 \pm  C_S^i (\bar C_V^i \mp \bar C_{IS}^i)   
 \nnl &
+ {\lambda_0 \over 3}  \bigg [    C_A^i  \bar C_P^i 
+    2 C_A^i  ( \bar C_{T(2) }^i  \pm  \alpha_T \bar C_T^i )  
+  2   C_T^i \bar C_M^i 
+   C_T^i  ( \bar C_{IT}^i \pm \alpha_A \bar C_A^i ) 
\bigg ] 
\bigg \} 
, \nnl 
\delta_1{\cal A}  = & 
{m_e \over  m_{\cal N} }  \re \bigg \{  
-  C_V^i  \bar C_P^i 
- C_A^i \bar C_{T(1)}^i
\mp    C_S^i (\bar C_A^i \pm \bar C_{IT}^i)  
\mp     C_T^i (\bar C_V^i \mp \bar C_{IS}^i)  
\nnl & 
\pm  \lambda_1  \bigg [ 
 C_A^i  ( \bar C_{T(2) }^i  \pm  \alpha_T \bar C_T^i ) 
+   C_T^i \bar C_M^i   \bigg ] 
+ \lambda_3 \bigg [ - C_A^i \bar C_{TQ}^i  +  C_T^i \bar C_Q^i  \bigg ]  
\bigg \} 
, \nnl 
\delta_1{\cal B}  = & 
{m_e \over  m_{\cal N} }\re \bigg \{  
- \bigg ( {E_0 \over E_e }  - 1 \bigg )  
C_V^i  \bar C_P^i  
\pm {E_0  \over  E_e   }  
C_V^i ( \bar C_T^i \pm \bar C_{T(2) }^i  ) 
-  \bigg ( {E_0 \over E_e }  - 1 \bigg ) C_A^i \bar C_{T(1) }^i
\nnl &
\mp  \bigg ( {E_0 \over E_e }  - 1 \bigg )   
C_S^i (\bar C_A^i \pm \bar C_{IT}^i) 
\pm     C_T^i (\bar C_V^i \mp \bar C_{IS}^i)  
\mp   \lambda_1 \bigg [ 
    C_A^i ( \bar C_{T(2) }^i  \pm  \alpha_T \bar C_T^i ) 
-   \bigg ( {E_0 \over E_e }  - 1 \bigg )   C_T^i \bar C_M^i   \bigg ] 
\nnl & 
+ \lambda_3    \bigg ( {E_0 \over E_e }  - 1 \bigg )    \bigg [ - C_A^i \bar C_{TQ}^i   + C_T^i \bar C_Q^i \bigg ]  
\bigg \} 
, \nnl 
\delta_1{\cal C}  = &
{m_e \over 2  m_{\cal N} } \re \bigg \{ 
C_S^i \bar C_Q^i  
+ \lambda_2 \bigg [  C_A^i  \bar C_P^i  
 -  C_A^i ( \bar C_{T(2) }^i  \pm  \alpha_T \bar C_T^i )
-   C_T^i \bar C_M^i    
+  C_T^i  ( \bar C_{IT}^i \pm \alpha_A \bar C_A^i )  \bigg ]  
\nnl & 
\pm 3 \lambda_4   \bigg [ - C_A^i \bar C_{TQ}^i  +    C_T^i \bar C_Q^i   \bigg ] 
\bigg \} 
, \nnl 
\delta_1{\cal D}  = & 
{m_e \over  m_{\cal N} } \im \bigg \{ 
\pm  C_V^i ( \bar C_T^i \pm \bar C_{T(2) }^i   )  
+  C_A^i \bar C_{T(1) }^i  
 + C_S^i \bar C_M^i
\pm   C_T^i (\bar C_V^i \mp \bar C_{IS}^i)   
\nnl &  
\mp { \lambda_1 \over   2  }  \bigg [ 
   C_A^i  \bar C_P^i    
 +    C_A^i ( \bar C_{T(2) }^i  \pm  \alpha_T \bar C_T^i )
+    C_T^i \bar C_M^i 
+     C_T^i  ( \bar C_{IT}^i \pm \alpha_A \bar C_A^i ) 
\bigg ] 
+  {\lambda_3  \over 2 } \bigg [ 
 - C_A^i \bar C_{TQ}^i + C_T^i \bar C_Q^i  \bigg ]
\bigg \} 
, \nnl  
\delta_1{\cal C}_4' = & 
{m_e \over  E_e } \re \bigg \{ 
- C_S^i \bar C_Q^i  
+ \lambda_2 \bigg [ 
-   C_A^i  \bar C_P^i   
+ C_A^i  ( \bar C_{T(2)}^i  \pm \alpha_T \bar C_T^i) 
+  C_T^i \bar C_M^i  
-  C_T^i  ( \bar C_{IT}^i \pm  \alpha_A \bar C_A^i) \bigg ] 
\nnl & 
\pm  3 \lambda_4 \bigg [ C_A^i \bar C_{TQ}^i  
- C_T^i \bar C_Q^i   \bigg ] 
\bigg \} 
,   \end{align}  
while the standard–exotic interference contributions to 
${\cal A}'$, ${\cal B}'$, ${\cal C}_3'$, ${\cal D}_3'$,  ${\cal D}_4'$, ${\cal E}_3'$, and ${\cal E}_4'$ vanish.
We remind that the upper (lower) sign refers to $\beta^-$ ($\beta^+$) decay, while $\lambda_i$ and  $\alpha_{A,T}$ are defined in \cref{tab:CORR_GroupTheoreticalLambda}. 

The first-recoil-order contributions to the form factors from interference between leading and recoil-suppressed pieces of the exotic currents take the form 
\begin{align}   
\label{eq:ERC_delta2}
\delta_2 {\cal N } = &  \re \bigg \{ 
\pm  {E_0 - 2 E_e + {m_e^2 \over E_e}  \over  m_{\cal N} }  C_S^i \bar C_{T(1)}^i 
\pm {\lambda_0 \over 3} \bigg [ 
{E_0 - 2 E_e + {m_e^2 \over E_e}  \over  m_{\cal N} }  C_T^i  \bar C_P^i    
-  {E_0 + {2 m_e^2 \over E_e}   \over m_{\cal N} } C_T^i ( \bar C_{T(2)}^i  \pm \alpha_T \bar C_T^i) 
\bigg ] 
\bigg \} 
, \nnl 
\delta_2 \aleph = & \re \bigg \{  
\pm  {2 E_e  - E_0  \over  m_{\cal N} }  C_S^i \bar C_{T(1)}^i 
\pm {\lambda_0 \over 3} \bigg [  
 {2 E_e  - E_0  \over  m_{\cal N} }  C_T^i  \bar C_P^i    
+   {E_0  \over m_{\cal N} } 
C_T^i ( \bar C_{T(2)}^i  \pm \alpha_T \bar C_T^i)  
\bigg ]  
\bigg \} 
, \nnl 
\delta_2 {\cal A}  = & \re \bigg \{  
\pm {E_0 - 2 E_e \over  m_{\cal N} }  
C_T^i \bar C_{T(1)}^i  
+ {\lambda_1 \over 2 } \bigg [  
{E_0  \over  m_{\cal N} }  C_T^i  \bar C_P^i   
+  {2 E_e - E_0  \over  m_{\cal N} } 
C_T^i ( \bar C_{T(2)}^i  \pm \alpha_T \bar C_T^i)  
\bigg ] 
\nnl & 
\mp \lambda_3  {14 E_e - E_0  \over 10 m_{\cal N} } C_T^i  \bar C_{TQ}^i
\bigg \} 
, \nnl 
\delta_2 {\cal B}  = & \re \bigg \{  
 {m_e^2 \over E_e m_{\cal N} } C_S^i
 (\bar C_T^i \pm \bar C_{T(2)}^i) 
\mp {E_0 - 2 E_e + {m_e^2 \over E_e}  \over  m_{\cal N} }  C_T^i \bar C_{T(1)}^i  
\nnl &  
+ {\lambda_1 \over 2 } \bigg [
-  {E_0 -  {m_e^2 \over E_e}  \over   m_{\cal N} }  C_T^i  \bar C_P^i   
+  {E_0 - 2 E_e - {m_e^2 \over E_e} \over  m_{\cal N} } C_T^i ( \bar C_{T(2)}^i  \pm \alpha_T \bar C_T^i)
\bigg ] 
\pm \lambda_3  {14 E_e - 13 E_0 - {m_e^2 \over E_e} \over 10 m_{\cal N} } C_T^i  \bar C_{TQ}^i  
\bigg \} 
, \nnl 
\delta_2 {\cal C}  = & 
 \re \bigg \{  
\pm  {E_0 - 2 E_e \over   2 m_{\cal N} } C_S^i  \bar C_{TQ}^i
+ \lambda_2  \bigg [ 
\pm  {2 E_e - E_0 \over   2 m_{\cal N} }   C_T^i  \bar C_P^i    
\pm  {E_0  \over 2 m_{\cal N} }  
C_T^i ( \bar C_{T(2)}^i  \pm \alpha_T \bar C_T^i)
\bigg ] 
- \lambda_4  {3 E_0  \over 2  m_{\cal N} } C_T^i  \bar C_{TQ}^i 
\bigg \} 
, \nnl 
\delta_2 {\cal D}  = & \im \bigg \{ 
\pm {2 E_e - E_0  \over  m_{\cal N} }  C_T^i \bar C_{T(1)}^i  
- {\lambda_1 \over 2 } \bigg [
 {E_0  \over   m_{\cal N} }  C_T^i  \bar C_P^i   
+  {2 E_e - E_0 \over m_{\cal N} }  
C_T^i \bar C_{T(2)}^i
\bigg ] 
\pm \lambda_3  {E_0- 2 E_e \over 2 m_{\cal N} } C_T^i  \bar C_{TQ}^i 
\bigg \} 
, \nnl 
\delta_2 {\cal A}' = &    \re \bigg \{  
C_S^i  \bar C_P^i   
+ C_S^i (\bar C_T^i \pm \bar C_{T(2)}^i)   
+ {\lambda_1 \over 2 } \bigg [ 
C_T^i  \bar C_P^i   
+ C_T^i ( \bar C_{T(2)}^i  \pm \alpha_T \bar C_T^i) 
\bigg ]  
\mp {3 \lambda_3 \over 10}  C_T^i  \bar C_{TQ}^i
\bigg \} 
, \nnl 
\delta_2 {\cal B}' = & \re \bigg \{  
C_S^i  \bar C_P^i  
- C_S^i (\bar C_T^i \pm \bar C_{T(2)}^i)  
+ {\lambda_1 \over 2 } \bigg [ 
-    C_T^i  \bar C_P^i   
+   C_T^i ( \bar C_{T(2)}^i  \pm \alpha_T \bar C_T^i)
\bigg ]  
\mp {3 \lambda_3 \over 10}  C_T^i  \bar C_{TQ}^i
\bigg \} 
, \nnl 
\delta_2 {\cal C}_3' = &  
 \re \bigg \{
\mp C_S^i  \bar C_{TQ}^i
+ \lambda_2  \bigg [ 
\pm  C_T^i  \bar C_P^i    
\pm  C_T^i ( \bar C_{T(2)}^i  \pm \alpha_T \bar C_T^i)
\bigg ] 
+ 3 \lambda_4 C_T^i  \bar C_{TQ}^i
\bigg \} 
, \nnl 
\delta_2 {\cal C}_4' = & 
 \re \bigg \{  
\pm C_S^i  \bar C_{TQ}^i 
+ \lambda_2  \bigg[ 
\mp  C_T^i  \bar C_P^i 
\pm  C_T^i ( \bar C_{T(2)}^i  \pm \alpha_T \bar C_T^i) 
\bigg ] 
+ 3 \lambda_4 C_T^i  \bar C_{TQ}^i 
\bigg \} 
, \nnl 
\delta_2 {\cal D}_3' = & 
  \im \bigg \{ 
\mp C_S^i  \bar C_{TQ}^i 
+ \lambda_2 \bigg [ 
\mp  C_T^i  \bar C_P^i 
\mp C_T^i  \bar C_{T(2)}^i
\bigg ] 
+ \lambda_4 C_T^i  \bar C_{TQ}^i
\bigg \} 
, \nnl 
\delta_2 {\cal D}_4' = &
 \im \bigg \{ 
 \mp C_S^i  \bar C_{TQ}^i 
+  \lambda_2  \bigg [ 
\mp  C_T^i  \bar C_P^i 
\pm  C_T^i \bar C_{T(2)}^i 
\bigg ] 
- \lambda_4 C_T^i  \bar C_{TQ}^i 
\bigg \} 
, \nnl 
\delta_2 {\cal E}_3' = & 
\re \bigg \{  
\mp \lambda_5 C_T^i  \bar C_{TQ}^i
\bigg \} 
, \nnl 
\delta_2 {\cal E}_4' = & \re \bigg \{   
\mp \lambda_5  C_T^i  \bar C_{TQ}^i 
\bigg \} 
.  \end{align}  
The contributions in \cref{eq:ERC_delta1,eq:ERC_delta2} should be added to those in \cref{eq:DIST_formfactors,eq:DIST_formfactors-primed} to obtain the complete characterization of the differential distribution in allowed beta decay at first recoil order.

\section{Gamow-Teller transitions} 
\label{app:GT}

In this appendix we briefly review  how the formalism spelled out in \cref{sec:ABD} can be modified to  describe allowed spin-changing transitions with 
$J_1 \equiv J = J_2 \pm 1$  and the same parity of the parent and daughter nuclei. 
These are referred to as (pure) Gamow-Teller transitions. 
As in the mixed Fermi-Gamow-Teller case, the $\beta$ decay amplitude is split into the leptonic and hadronic current parts, cf. \cref{eq:FORM_Mbeta-current}.
For $\beta^-$ decays with $J = J_2 + 1$,  the most general  hadronic currents to subleading order in recoil with the correct little group transformation properties and consistent with the parity constraint in \cref{eq:FORM_Pconstraint} can be written in the spinorial form as 
\begin{align}
\label{eq:GT_currents-spinorial-descending}
 & {\bf J = J_2 + 1 } 
\nnl 
j_{V- }^\mu = & 
i {C_M^i \over m_{\cal N}^2}    
\epsilon^{\mu \nu \alpha \beta } q_\nu p_1^\alpha 
(\chi_1 \sigma^\beta \tilde \chi_1)     
  { \big [ \chi_2 \chi_1 +  \tilde \chi_2 \tilde \chi_1  \big ]^{2 J_2} \over (2 m_{\cal N})^{2 J_2} }    
 \nnl & 
-  {C_Q^i \over  m_{\cal N}^2 } 
 ( \chi_1  \sigma^{ \{ \nu }  \tilde \chi_1) q_\nu
 \big [ \chi_2  \sigma^{\mu \} }   \tilde \chi_1 -  \tilde \chi_2  \bar \sigma^{\mu \} } \chi_1 \big ]  
  { \big [ \chi_2 \chi_1 +  \tilde \chi_2 \tilde \chi_1  \big ]^{2 J_2-1}
  \over (2 m_{\cal N} )^{2 J_2 - 1 } } 
+ \cO(q^2)
,\nnl 
j_{A-}^\mu = & 
\bigg \{   2 C_{A}^i ( \chi_1  \sigma^\mu \tilde \chi_1) 
+ { C_{IT}^i  \over  m_{\cal N}^2 } 
p_1^\mu (\chi_1 q \sigma \tilde \chi_1)     
\bigg \} 
  { \big [ \chi_2 \chi_1 +  \tilde \chi_2 \tilde \chi_1  \big ]^{2 J_2} \over (2 m_{\cal N})^{2 J_2} }    
+ \cO(q^2) 
,\nnl  
j_{S-} = & \cO(q^2)  
,\nnl  
j_{P-} = &  { C_P^i \over  m_{\cal N} }  (\chi_1 q \sigma \tilde \chi_1)  
  { \big [ \chi_2 \chi_1 +  \tilde \chi_2 \tilde \chi_1  \big ]^{2 J_2} \over (2 m_{\cal N})^{2 J_2} }   
  + \cO(q^2) 
,\nnl  
j_{T-}^{\mu\nu} = & 
-i \bigg \{ 
 {2 C_T^i \over  m_{\cal N} } 
 p_1^{[\mu} (\chi_1 \sigma^{\nu]} \tilde \chi_1)  
+  { C_{T(2)}^i \over  m_{\cal N} } 
(\chi_1 \sigma^{[\mu} \tilde \chi_1) q^{\nu]}  
  \bigg \}  
  {   \big [ \chi_2 \chi_1  + \tilde \chi_2 \tilde \chi_1  \big ]^{2J_2}   \over (2 m_{\cal N})^{2J_2} } 
    \nnl +  & 
  i { C_{TQ}^i \over  2 m_{\cal N}^3 }  
 q_\alpha (\chi_1\sigma^{\{ \alpha} \tilde \chi_1)  
 \bigg ( \big [ \chi_2 \sigma^{\mu \} } \tilde \chi_1 - \tilde \chi_2 \bar \sigma^{\mu \} }  \chi_1 \big ]    p_1^{\nu}
 -  (\mu \leftrightarrow \nu)  \bigg )
   {   \big [ \chi_2 \chi_1  + \tilde \chi_2 \tilde \chi_1  \big ]^{2J_2-1}   \over (2 m_{\cal N})^{2J_2-1} }  
  + \cO(q^2)   
,  \end{align}
with the understanding that the $C_Q^i$ and $C_{TQ}^i$ structures are absent for $J=1$. 
For $J = J_2 - 1$, on the other hand, the currents with correct little group and parity transformation properties can be obtained from \cref{eq:GT_currents-spinorial-descending} by  $1 \leftrightarrow 2$.  
Adjusting the signs for future convenience,  we obtain the currents
 \begin{align}
\label{eq:GT_currents-spinorial-ascending}
& {\bf J = J_2 - 1 } 
\nnl 
j_{V- }^\mu = & 
 - i  {C_M^i   \over m_{\cal N}^2} \epsilon^{\mu \nu \alpha \beta }   q_\nu p_1^\alpha  
(\chi_2 \sigma^\beta \tilde \chi_2)      
  { \big [ \chi_2 \chi_1 +  \tilde \chi_2 \tilde \chi_1  \big ]^{2 J} \over (2 m_{\cal N})^{2 J} }    
 \nnl & 
 +   { C_Q^i  \over m_{\cal N}^2} ( \chi_2  \sigma^{ \{ \nu }  \tilde \chi_2) q_\nu
 \big [ \chi_2  \sigma^{\mu \} }   \tilde \chi_1 -  \tilde \chi_2  \bar \sigma^{\mu \} } \chi_1 \big ]  
  { \big [ \chi_2 \chi_1 +  \tilde \chi_2 \tilde \chi_1  \big ]^{2 J-1}
  \over (2 m_{\cal N} )^{2 J - 1 } } 
+ \cO(q^2)
,\nnl 
j_{A-}^\mu = & 
  - 2  C_{A}^i ( \chi_2  \sigma^\mu \tilde \chi_2)  
{ \big [ \chi_2 \chi_1 +  \tilde \chi_2 \tilde \chi_1  \big ]^{2 J} \over (2 m_{\cal N})^{2 J} }  
-  { C_{IT}^i   \over m_{\cal N}^2}  (\chi_2 q \sigma \tilde \chi_2)    p_1^\mu  
  { \big [ \chi_2 \chi_1 +  \tilde \chi_2 \tilde \chi_1  \big ]^{2 J} \over (2 m_{\cal N})^{2 J} }    
+ \cO(q^2) 
,\nnl  
j_{S-} = & \cO(q^2)  
,\nnl  
j_{P-} = &  - {C_P^i \over   m_{\cal N} } (\chi_2 q \sigma \tilde \chi_2)  
  { \big [ \chi_2 \chi_1 +  \tilde \chi_2 \tilde \chi_1  \big ]^{2 J} \over (2 m_{\cal N})^{2 J} }  
  + \cO(q^2) 
,\nnl  
j_{T-}^{\mu\nu} = & 
  i \bigg \{  { 2 C_T^i  \over  m_{\cal N} }p_1^{[\mu}(\chi_2 \sigma^{\nu ] } \tilde \chi_2)   
+   { C_{T(2)}^i \over m_{\cal N}  }    (\chi_2 \sigma^{[\mu} \tilde \chi_2) q^{\nu]}  
\bigg \} 
{ \big [ \chi_2 \chi_1 +  \tilde \chi_2 \tilde \chi_1  \big ]^{2 J}      \over (2 m_{\cal N} )^{2 J} } 
 \nnl -  & 
  i { C_{TQ}^i \over  2 m_{\cal N}^3 }  
 q_\alpha (\chi_2\sigma^{\{ \alpha} \tilde \chi_2)  
 \bigg ( \big [ \chi_2 \sigma^{\mu \} } \tilde \chi_1 - \tilde \chi_2 \bar \sigma^{\mu \} }  \chi_1 \big ]    p_1^{\nu}
 -  (\mu \leftrightarrow \nu)  \bigg )
   {   \big [ \chi_2 \chi_1  + \tilde \chi_2 \tilde \chi_1  \big ]^{2J-1}   \over (2 m_{\cal N})^{2J-1} }
+ \cO(q^2)  
,  \end{align}
with the understanding that the $C_Q^i$ and $C_{TQ}^i$ structures are absent for $J=0$.

Expanding the spinor bilinears in the recoil momentum up to $\cO(q^2)$ corrections,  both \cref{eq:GT_currents-spinorial-descending,eq:GT_currents-spinorial-ascending}   lead to
\begin{align}
\label{eq:GT_currents}
j_{V - }^\mu = & 
 i { C_M^i  \over m_{\cal N} }   \epsilon^{\mu \nu \alpha \beta }   q_\nu  p_1^\alpha    \Sigma^\beta 
+   C_Q^i q_\nu   \Sigma^{\mu\nu} 
,\nnl 
j_{A - }^\mu = & 
 2  m_{\cal N} C_{A}^i     \Sigma^\mu 
+   {C_{IT}^i + \alpha_A C_A^i  \over  m_{\cal N} }   p_1^\mu   q_\nu   \Sigma^\nu    
,\nnl  
j_{S - } = & 0  
,\nnl  
j_{P - } = &  C_P^i   q_\nu   \Sigma^\nu    
,\nnl  
j_{T- }^{\mu\nu} = & 
-  i C_T^i \big [   p_1^\mu  \Sigma^\nu  -  p_1^\nu  \Sigma^\mu   \big ]   
+   i { C_{T(2)}^i  \over 2   }  \big [   q^\mu  \Sigma^\nu  -  q^\nu  \Sigma^\mu   \big ] 
+ i {C_{TQ}^i \over 2 m_{\cal N}} \big [ p_1^{\mu}  \Sigma^{\nu \alpha} -  p_1^{\nu}  \Sigma^{\mu \alpha}   \big ]  q_\alpha  
,  \end{align}
where $\alpha_A$ is defined in \cref{tab:CORR_GroupTheoreticalLambda}, and   
\begin{itemize}
    \item ${\bf J = J_2 + 1 }$
 \begin{align} 
\, [\Sigma^\mu]^{I_1  \dots I_{2J} }_{O_3  \dots O_{2J } } \equiv & 
{ (\chi_1^{I_1 } \sigma^\mu \tilde \chi_1^{I_2 })    \over  m_{\cal N} }
 \delta^{I_3}_{O_3} \dots   \delta^{I_{2J}}_{O_{2J} }  
 , \nnl 
 \, [\Sigma^{\mu\nu}]^{I_1  \dots I_{2J} }_{O_3 \dots O_{2J } } \equiv & 
{ (\chi_1^{I_1 } \sigma^{\{ \mu } \tilde \chi_1^{I_2 })   
(\chi_{1\, O_3 } \sigma^{\nu \} } \tilde \chi_1^{I_3 } + \tilde \chi_{1\, O_3 } \bar \sigma^{\nu \} } \chi_1^{I_3 })    \over  m_{\cal N}^2 }
 \delta^{I_4}_{O_4} \dots   \delta^{I_{2J}}_{O_{2J} }  
,  \end{align}
\item ${\bf J = J_2 - 1 }$
\begin{align} 
\, [\Sigma^\mu]^{I_1  \dots I_{2J} }_{O_1  \dots O_{2J+2 } } \equiv & 
{1 \over   m_{\cal N} } (\chi_{1\, O_{2J+1} } \sigma^\mu \tilde \chi_{1\, O_{2J+2 } })  
 \delta^{I_1}_{O_1} \dots   \delta^{I_{2J}}_{O_{2J} }  
 , \nnl 
 \, [\Sigma^{\mu \nu}]^{I_1  \dots I_{2J} }_{O_1 \dots O_{2J +2} } \equiv & 
{1 \over  m_{\cal N}^2  } (\chi_{1\, O_{2J+1} } \sigma^{\{ \mu }  \tilde \chi_{1\, O_{2J+2} })
(\chi_{1\, O_1 } \sigma^{\nu \} } \tilde \chi_1^{I_1 } + \tilde \chi_{1 \, O_1 } \bar \sigma^{\nu \} } \chi_1^{I_1 }  )   
 \delta^{I_2}_{O_2} \dots   \delta^{I_{2J}}_{O_{2J} }  
. \end{align}
\end{itemize}

$\Sigma^\mu$ and $\Sigma^{\mu\nu}$ are analogous to the spin vector and tensor defined in \cref{eq:SI_spinVector,eq:SI_spinTensor}, but with the little-group index structure appropriate for spin changing by one unit. 
We also define 
$[\bar \Sigma^\mu]_{I_1  \dots I_{2J} }^{O_3  \dots O_{2J } }\equiv \big ( [\Sigma^\mu]^{I_1  \dots I_{2J} }_{O_3  \dots O_{2J }} \big)^* $ for $J = J_2 +1$,
$[\bar \Sigma^\mu]_{I_1  \dots I_{2J} }^{O_1  \dots O_{2J +2 } }\equiv \big ( [\Sigma^\mu]^{I_1  \dots I_{2J} }_{O_1 \dots O_{2J+2 }} \big)^* $ for $J = J_2 -1$, and likewise for $\bar \Sigma^{\mu \nu}$.
We have the identities 
\begin{itemize}
    \item ${\bf J = J_2 + 1 }$ 
\begin{align}   
 \label{eq:GT_SigmaMuBarSigmaNu-descending}  
 \Sigma^\mu  \bar \Sigma^\nu= &  
- {2 \over 3}  \Pi^{\mu \nu}  \mathbbm{1} 
+ { i  \over J m_{\cal N } } \epsilon^{\mu \nu \alpha \beta} p_1{}_\alpha S_\beta 
- {1 \over J ( J - 1/2) } S^{\mu \nu} 
, \quad J \geq 1 
, \nnl 
\Sigma^{\mu} \bar \Sigma^{\nu \rho} = & 
   {1 \over 5 J} \bigg [ 3 \Pi^{\mu  \nu} S^\rho  +3  \Pi^{\mu  \rho} S^\nu
 - 2   \Pi^{\nu\rho} S^\mu  \bigg ] 
 -  { i  \over  J(J-1/2) m_{\cal N } }
\bigg [  \epsilon^{\mu \nu \alpha \beta}  p_1{}_\alpha  S_{\beta}{}^\rho
+ \epsilon^{\mu \rho \alpha \beta}  p_1{}_\alpha    S_{\beta}{}^\nu \bigg ] 
\nnl & 
  + {2   \over J(J-1/2)(J-1) } S^{\mu \nu \rho }
  , \quad J \geq 3/2 
,  \end{align} 
\item ${\bf J = J_2 - 1 } $ 
\begin{align}   
 \label{eq:GT_SigmamuBarSigmanu-ascending} 
  \Sigma^\mu  \bar \Sigma^\nu = & 
  - {2 J +3 \over 2J +1 }\bigg \{ 
 {2  \over 3 } \Pi^{\mu \nu} \mathbbm{1} 
+  { i \over (J+1) m_{\cal N } } \epsilon^{\mu \nu \alpha \beta} p_1{}_\alpha S_\beta 
+  {2 \over (J+1) ( 2 J + 3) } S^{\mu \nu}  \bigg \} 
, \nnl 
\Sigma^{\mu} \bar \Sigma^{\nu \rho} = & 
{1 \over J (J+1/2) (J+1) } \bigg \{ 
{ (J+3/2) (J+2) \over 5} \bigg [3 \Pi^{\mu  \nu} S^\rho  +3  \Pi^{\mu  \rho} S^\nu  
-  2    \Pi^{\nu\rho} S^\mu  \bigg ] 
\nnl &  
 +  { i (J+2)  \over  m_{\cal N } }
\bigg [  \epsilon^{\mu \nu \alpha \beta}  p_1{}_\alpha  S_\beta{}^\rho
+ \epsilon^{\mu \rho \alpha \beta}  p_1{}_\alpha    S_\beta{}^\nu \bigg ] 
  + 2  S^{\mu \nu \rho }
\bigg \} 
  , \quad J \geq 1/2  
, \end{align}  
\end{itemize} 
where $\Pi^{\mu \nu} \equiv \eta^{\mu \nu} - {p_1^\mu p_1^\nu \over m_{\cal N}^2}$, and  $S^{\mu \nu \rho }$ is defined in \cref{eq:SI_spinTensor}. 

In these formulas, the left-hand sides implicitly involve summing over the little group indices of the daughter nucleus. 
In \cref{eq:GT_SigmamuBarSigmanu-ascending} it is implicit that $S^\mu$ should be omitted for $J=0$, 
$S^{\mu\nu}$ should be omitted for $J \leq 1/2$, 
and $S^{\mu\nu \rho}$ should be omitted for $J \leq 1$.  

For $\beta^+$ decay our convention is that we replace  in \cref{eq:GT_currents-spinorial-descending,eq:GT_currents-spinorial-ascending,eq:GT_currents}
$C_{A,T,M,P}^i \to \bar C_{A,T,M,P}^i$, 
and 
$C_{Q,IT,T(2),TQ}^i \to - \bar C_{Q,IT,T(2),TQ}^i$.

\bibliographystyle{JHEP} 
\bibliography{ForbiddenPlanet}

@article{Sharma:2009hg,
    author = "Sharma, Neetika and Dahiya, Harleen and Chatley, P. K. and Gupta, Manmohan",
    title = "{Weak vector and axial-vector form factors in the chiral constituent quark model with configuration mixing}",
    eprint = "0904.2246",
    archivePrefix = "arXiv",
    primaryClass = "hep-ph",
    doi = "10.1103/PhysRevD.79.077503",
    journal = "Phys. Rev. D",
    volume = "79",
    pages = "077503",
    year = "2009"
}

@article{Shiomi:1996np,
    author = "Shiomi, H.",
    title = "{Second class current in QCD sum rules}",
    eprint = "hep-ph/9601329",
    archivePrefix = "arXiv",
    reportNumber = "UTHEP-329",
    doi = "10.1016/0375-9474(96)80003-A",
    journal = "Nucl. Phys. A",
    volume = "603",
    pages = "281--302",
    year = "1996"
}

@article{Dubbers:2021wqv,
    author = {Dubbers, Dirk and M{\"a}rkisch, Bastian},
    title = "{Precise Measurements of the Decay of Free Neutrons}",
    eprint = "2106.02345",
    archivePrefix = "arXiv",
    primaryClass = "nucl-ex",
    doi = "10.1146/annurev-nucl-102419-043156",
    journal = "Ann. Rev. Nucl. Part. Sci.",
    volume = "71",
    pages = "139--163",
    year = "2021"
}

@article{Hayen:2017pwg,
    author = "Hayen, Leendert and Severijns, Nathal and Bodek, Kazimierz and Rozpedzik, Dagmara and Mougeot, Xavier",
    title = "{High precision analytical description of the allowed $\beta$ spectrum shape}",
    eprint = "1709.07530",
    archivePrefix = "arXiv",
    primaryClass = "nucl-th",
    doi = "10.1103/RevModPhys.90.015008",
    journal = "Rev. Mod. Phys.",
    volume = "90",
    number = "1",
    pages = "015008",
    year = "2018"
}

@article{Vanlangendonck:2022dud,
    author = {Vanlangendonck, S. and Severijns, N. and Hayen, L. and Gl{\"u}ck, F.},
    title = "{Influence of the recoil-order and radiative correction on the {\ensuremath{\beta}} decay correlation coefficients in mirror decays}",
    eprint = "2203.16936",
    archivePrefix = "arXiv",
    primaryClass = "nucl-ex",
    doi = "10.1103/PhysRevC.106.015506",
    journal = "Phys. Rev. C",
    volume = "106",
    number = "1",
    pages = "015506",
    year = "2022"
}

@unpublished{inprogress,
  author = {Cruz-Alzaga, Sergio and Falkowski, Adam and  Gonz{\'a}lez-Alonso, Mart{\'\i}n and Palavri{\'c}, Ajdin},
  title  = {\em work in progress},
  note   = {Work in progress},
  year   = {2026}
}

@article{Dawson:2021ofa,
    author = "Dawson, Sally and Giardino, Pier Paolo",
    title = "{New physics through Drell-Yan standard model EFT measurements at NLO}",
    eprint = "2105.05852",
    archivePrefix = "arXiv",
    primaryClass = "hep-ph",
    doi = "10.1103/PhysRevD.104.073004",
    journal = "Phys. Rev. D",
    volume = "104",
    number = "7",
    pages = "073004",
    year = "2021"
}

@article{deBlas:2013qqa,
    author = "de Blas, Jorge and Chala, Mikael and Santiago, Jose",
    title = "{Global Constraints on Lepton-Quark Contact Interactions}",
    eprint = "1307.5068",
    archivePrefix = "arXiv",
    primaryClass = "hep-ph",
    doi = "10.1103/PhysRevD.88.095011",
    journal = "Phys. Rev. D",
    volume = "88",
    pages = "095011",
    year = "2013"
}

@article{Gorchtein:2023srs,
    author = "Gorchtein, Mikhail and Seng, Chien-Yeah",
    title = "{The Standard Model Theory of Neutron Beta Decay}",
    eprint = "2307.01145",
    archivePrefix = "arXiv",
    primaryClass = "hep-ph",
    doi = "10.3390/universe9090422",
    journal = "Universe",
    volume = "9",
    number = "9",
    pages = "422",
    year = "2023"
}

@article{Gell-Mann:1958sai,
    author = "Gell-Mann, Murray",
    title = "{Test of the Nature of the Vector Interaction in beta Decay}",
    doi = "10.1103/PhysRev.111.362",
    journal = "Phys. Rev.",
    volume = "111",
    pages = "362--365",
    year = "1958"
}

@article{Calaprice:1976jbi,
    author = "Calaprice, Frank P. and Holstein, Barry R.",
    title = "{Weak magnetism and the beta spectra of 12 B and 12 N}",
    doi = "10.1016/0375-9474(76)90593-5",
    journal = "Nucl. Phys. A",
    volume = "273",
    pages = "301--325",
    year = "1976"
}

@article{Cirigliano:2022hob,
    author = "Cirigliano, Vincenzo and de Vries, Jordy and Hayen, Leendert and Mereghetti, Emanuele and Walker-Loud, Andr{\'e}",
    title = "{Pion-Induced Radiative Corrections to Neutron {\ensuremath{\beta}} Decay}",
    eprint = "2202.10439",
    archivePrefix = "arXiv",
    primaryClass = "nucl-th",
    reportNumber = "LA-UR-21-31960, INT-PUB-22-005, LA-UR-21-31960; INT-PUB-22-005",
    doi = "10.1103/PhysRevLett.129.121801",
    journal = "Phys. Rev. Lett.",
    volume = "129",
    number = "12",
    pages = "121801",
    year = "2022"
}

@article{FlavourLatticeAveragingGroupFLAG:2024oxs,
    author = "Aoki, Y. and others",
    collaboration = "Flavour Lattice Averaging Group (FLAG)",
    title = "{FLAG review 2024}",
    eprint = "2411.04268",
    archivePrefix = "arXiv",
    primaryClass = "hep-lat",
    reportNumber = "CERN-TH-2024-192, FERMILAB-PUB-24-0785-T",
    doi = "10.1103/nfzp-p5dn",
    journal = "Phys. Rev. D",
    volume = "113",
    number = "1",
    pages = "014508",
    year = "2026"
}

@article{Weinberg:1958ut,
    author = "Weinberg, Steven",
    title = "{Charge symmetry of weak interactions}",
    doi = "10.1103/PhysRev.112.1375",
    journal = "Phys. Rev.",
    volume = "112",
    pages = "1375--1379",
    year = "1958"
}

@article{Severijns:2006dr,
    author = "Severijns, Nathal and Beck, Marcus and Naviliat-Cuncic, Oscar",
    title = "{Tests of the standard electroweak model in beta decay}",
    eprint = "nucl-ex/0605029",
    archivePrefix = "arXiv",
    doi = "10.1103/RevModPhys.78.991",
    journal = "Rev. Mod. Phys.",
    volume = "78",
    pages = "991--1040",
    year = "2006"
}

@article{Holstein:2008sx,
    author = "Holstein, Barry R. and Ross, Andreas",
    title = "{Spin Effects in Long Range Gravitational Scattering}",
    eprint = "0802.0716",
    archivePrefix = "arXiv",
    primaryClass = "hep-ph",
    month = "2",
    year = "2008"
}

@article{Gell-Mann:1962yej,
    author = "Gell-Mann, Murray",
    title = "{Symmetries of baryons and mesons}",
    doi = "10.1103/PhysRev.125.1067",
    journal = "Phys. Rev.",
    volume = "125",
    pages = "1067--1084",
    year = "1962"
}

@article{Holstein:1974zf,
      author         = "Holstein, Barry R.",
      title          = "{Recoil Effects in Allowed beta Decay: The Elementary
                        Particle Approach}",
      journal        = "Rev. Mod. Phys.",
      volume         = "46",
      year           = "1974",
      pages          = "789",
      doi            = "10.1103/RevModPhys.46.789",
      note           = "[Erratum: Rev. Mod. Phys.48,673(1976)]",
      reportNumber   = "Print-74-0849 (MASS.U.,AMHERST)",
      SLACcitation   = "%%CITATION = RMPHA,46,789;%%"
}

@unpublished{Melconian:RecoilOrder,
    author = {Melconian, D.},
    title  = {Implementing recoil-order corrections (unpublished)},
    note   = {Unpublished TRINAT internal report},
    url    = {https://trinat.triumf.ca/documents/old-report-on-recoil-order-corrections},
    urldate = {2026-10-01}
}

@article{Cangemi:2023ysz,
    author = "Cangemi, Lucile and Chiodaroli, Marco and Johansson, Henrik and Ochirov, Alexander and Pichini, Paolo and Skvortsov, Evgeny",
    title = "{From higher-spin gauge interactions to Compton amplitudes for root-Kerr}",
    eprint = "2311.14668",
    archivePrefix = "arXiv",
    primaryClass = "hep-th",
    reportNumber = "UUITP-35/23",
    doi = "10.1007/JHEP09(2024)196",
    journal = "JHEP",
    volume = "09",
    pages = "196",
    year = "2024"
}

@article{Aoude:2020onz,
    author = "Aoude, Rafael and Haddad, Kays and Helset, Andreas",
    title = "{On-shell heavy particle effective theories}",
    eprint = "2001.09164",
    archivePrefix = "arXiv",
    primaryClass = "hep-th",
    reportNumber = "SAGEX-20-02-E, MITP/20-002",
    doi = "10.1007/JHEP05(2020)051",
    journal = "JHEP",
    volume = "05",
    pages = "051",
    year = "2020"
}

@article{Alviani:2024fby,
    author = "Alviani, Edoardo and Falkowski, Adam",
    title = "{On the Coulomb corrections in nuclear beta decay}",
    eprint = "2412.17702",
    archivePrefix = "arXiv",
    primaryClass = "hep-ph",
    doi = "10.1007/JHEP07(2025)059",
    journal = "JHEP",
    volume = "07",
    pages = "059",
    year = "2025"
}

@article{Falkowski:2021vdg,
    author = "Falkowski, Adam and Gonz{\'a}lez-Alonso, Mart{\'\i}n and Palavri{\'c}, Ajdin and Rodr{\'\i}guez-S{\'a}nchez, Antonio",
    title = "{Constraints on subleading interactions in beta decay Lagrangian}",
    eprint = "2112.07688",
    archivePrefix = "arXiv",
    primaryClass = "hep-ph",
    doi = "10.1007/JHEP02(2024)091",
    journal = "JHEP",
    volume = "02",
    pages = "091",
    year = "2024"
}

@article{Cirigliano:2012ab,
    author = "Cirigliano, Vincenzo and Gonzalez-Alonso, Martin and Graesser, Michael L.",
    title = "{Non-standard Charged Current Interactions: beta decays versus the LHC}",
    eprint = "1210.4553",
    archivePrefix = "arXiv",
    primaryClass = "hep-ph",
    reportNumber = "LA-UR-12-24810, NPAC-12-14",
    doi = "10.1007/JHEP02(2013)046",
    journal = "JHEP",
    volume = "02",
    pages = "046",
    year = "2013"
}

@article{Herczeg:2001vk,
    author = "Herczeg, P.",
    title = "{Beta decay beyond the standard model}",
    doi = "10.1016/S0146-6410(01)00149-1",
    journal = "Prog. Part. Nucl. Phys.",
    volume = "46",
    pages = "413--457",
    year = "2001"
}

@article{Arkani-Hamed:2017jhn,
    author = "Arkani-Hamed, Nima and Huang, Tzu-Chen and Huang, Yu-tin",
    title = "{Scattering amplitudes for all masses and spins}",
    eprint = "1709.04891",
    archivePrefix = "arXiv",
    primaryClass = "hep-th",
    reportNumber = "NCTS-TH/1714, NCTS-TH-1714",
    doi = "10.1007/JHEP11(2021)070",
    journal = "JHEP",
    volume = "11",
    pages = "070",
    year = "2021"
}

@article{Fermi:1934hr,
    author = "Fermi, E.",
    title = "{An attempt of a theory of beta radiation. 1.}",
    reportNumber = "UCRL-TRANS-726",
    doi = "10.1007/BF01351864",
    journal = "Z. Phys.",
    volume = "88",
    pages = "161--177",
    year = "1934"
}

@article{Falkowski:2020pma,
    author = "Falkowski, Adam and Gonz\'alez-Alonso, Mart\'\i{}n and Naviliat-Cuncic, Oscar",
    title = "{Comprehensive analysis of beta decays within and beyond the Standard Model}",
    eprint = "2010.13797",
    archivePrefix = "arXiv",
    primaryClass = "hep-ph",
    reportNumber = "IFIC/20-49, FTUV/20-1027",
    doi = "10.1007/JHEP04(2021)126",
    journal = "JHEP",
    volume = "04",
    pages = "126",
    year = "2021"
}

@article{Bhattacharya:2011qm,
    author = "Bhattacharya, Tanmoy and Cirigliano, Vincenzo and Cohen, Saul D. and Filipuzzi, Alberto and Gonzalez-Alonso, Martin and Graesser, Michael L. and Gupta, Rajan and Lin, Huey-Wen",
    title = "{Probing Novel Scalar and Tensor Interactions from (Ultra)Cold Neutrons to the LHC}",
    eprint = "1110.6448",
    archivePrefix = "arXiv",
    primaryClass = "hep-ph",
    reportNumber = "LA-UR-11-11460, NT@UW-11-16, IFIC-11-57, FTUV-11-1007, NPAC-11-14",
    doi = "10.1103/PhysRevD.85.054512",
    journal = "Phys. Rev. D",
    volume = "85",
    pages = "054512",
    year = "2012"
}

@article{Hardy:2020qwl,
    author = "Hardy, J.C. and Towner, I.S.",
    title = "{Superallowed $0^+\to 0^+$ nuclear \ensuremath{\beta} decays: 2020 critical survey, with implications for Vud and CKM unitarity}",
    doi = "10.1103/PhysRevC.102.045501",
    journal = "Phys. Rev. C",
    volume = "102",
    number = "4",
    pages = "045501",
    year = "2020"
}

@article{Severijns:2008ep,
      author         = "Severijns, N. and Tandecki, M. and Phalet, T. and Towner,
                        I. S.",
      title          = "{Ft values of the T = 1/2 mirror beta transitions}",
      journal        = "Phys. Rev.",
      volume         = "C78",
      year           = "2008",
      pages          = "055501",
      doi            = "10.1103/PhysRevC.78.055501",
      eprint         = "0807.2201",
      archivePrefix  = "arXiv",
      primaryClass   = "nucl-ex",
      SLACcitation   = "%%CITATION = ARXIV:0807.2201;%%"
      }

@article{Lee:1956qn,
      author         = "Lee, T. D. and Yang, Chen-Ning",
      title          = "{Question of Parity Conservation in Weak Interactions}",
      journal        = "Phys. Rev.",
      volume         = "104",
      year           = "1956",
      pages          = "254-258",
      doi            = "10.1103/PhysRev.104.254",
      SLACcitation   = "%%CITATION = PHRVA,104,254;%%"
}

@article{Ademollo:1964sr,
      author         = "Ademollo, M. and Gatto, Raoul",
      title          = "{Nonrenormalization Theorem for the Strangeness Violating
                        Vector Currents}",
      journal        = "Phys. Rev. Lett.",
      volume         = "13",
      year           = "1964",
      pages          = "264-265",
      doi            = "10.1103/PhysRevLett.13.264",
      SLACcitation   = "%%CITATION = PRLTA,13,264;%%"
}

@article{Gonzalez-Alonso:2018omy,
      author         = "Gonzalez-Alonso, Martin and Naviliat-Cuncic, Oscar and
                        Severijns, Nathal",
      title          = "{New physics searches in nuclear and neutron $\beta$ decay}",
      journal        = "Prog. Part. Nucl. Phys.",
      volume         = "104",
      year           = "2019",
      pages          = "165-223",
      doi            = "10.1016/j.ppnp.2018.08.002",
      eprint         = "1803.08732",
      archivePrefix  = "arXiv",
      primaryClass   = "hep-ph",
      reportNumber   = "CERN-TH-2018-050",
      SLACcitation   = "%%CITATION = ARXIV:1803.08732;%%"
}

@article{Jackson:1957zz,
      author         = "Jackson, J. D. and Treiman, S. B. and Wyld, H. W.",
      title          = "{Possible tests of time reversal invariance in Beta
                        decay}",
      journal        = "Phys. Rev.",
      volume         = "106",
      year           = "1957",
      pages          = "517-521",
      doi            = "10.1103/PhysRev.106.517",
      SLACcitation   = "%%CITATION = PHRVA,106,517;%%"
}

@article{Donoghue:1990ti,
      author         = "Donoghue, John F. and Wyler, D.",
      title          = "{Isospin Breaking and the Precise Determination of $V (u
                        d$)}",
      journal        = "Phys. Lett.",
      volume         = "B241",
      year           = "1990",
      pages          = "243-248",
      doi            = "10.1016/0370-2693(90)91287-L",
      reportNumber   = "CERN-TH-5634/90",
      SLACcitation   = "%%CITATION = PHLTA,B241,243;%%"
}

@article{GellMann:1964tf,
      author         = "Gell-Mann, Murray",
      title          = "{The Symmetry group of vector and axial vector currents}",
      journal        = "Physics",
      volume         = "1",
      pages          = "63-75",
      year           = "1964",
      SLACcitation   = "%%CITATION = PYCSA,1,63;%%",
}

@ARTICLE{Abele:2008zz,
  author = {Abele, Hartmut},
  title = {{The neutron. Its properties and basic interactions}},
  journal = {Prog. Part. Nucl. Phys.},
  year = {2008},
  volume = {60},
  pages = {1-81},
  doi = {10.1016/j.ppnp.2007.05.002},
  slaccitation = {%%CITATION = PPNPD,60,1;%%}
}

@article{Dreiner:2008tw,
      author         = "Dreiner, Herbi K. and Haber, Howard E. and Martin,
                        Stephen P.",
      title          = "{Two-component spinor techniques and Feynman rules for
                        quantum field theory and supersymmetry}",
      journal        = "Phys.Rept.",
      volume         = "494",
      pages          = "1-196",
      doi            = "10.1016/j.physrep.2010.05.002",
      year           = "2010",
      eprint         = "0812.1594",
      archivePrefix  = "arXiv",
      primaryClass   = "hep-ph",
      reportNumber   = "BN-TH-2008-12, SCIPP-08-08, FERMILAB-PUB-09-855-T",
      SLACcitation   = "%%CITATION = ARXIV:0812.1594;%%",
}

@article{Falkowski:2017pss,
      author         = "Falkowski, Adam and Gonzalez-Alonso, Martin and
                        Mimouni, Kin",
      title          = "{Compilation of low-energy constraints on 4-fermion
                        operators in the SMEFT}",
      journal        = "JHEP",
      volume         = "08",
      year           = "2017",
      pages          = "123",
      doi            = "10.1007/JHEP08(2017)123",
      eprint         = "1706.03783",
      archivePrefix  = "arXiv",
      primaryClass   = "hep-ph",
      SLACcitation   = "%%CITATION = ARXIV:1706.03783;%%"
}

@article{Gonzalez-Alonso:2013ura,
      author         = "Gonzalez-Alonso, Martin and Martin Camalich, Jorge",
      title          = "{Isospin breaking in the nucleon mass and the sensitivity
                        of $\beta$ decays to new physics}",
      journal        = "Phys. Rev. Lett.",
      volume         = "112",
      year           = "2014",
      number         = "4",
      pages          = "042501",
      doi            = "10.1103/PhysRevLett.112.042501",
      eprint         = "1309.4434",
      archivePrefix  = "arXiv",
      primaryClass   = "hep-ph",
      SLACcitation   = "%%CITATION = ARXIV:1309.4434;%%"
}

@article{Severijns:2021lsp,
    author = "Severijns, N. and Hayen, L. and De Leebeeck, V. and Vanlangendonck, S. and Bodek, K. and Rozpedzik, D. and Towner, I. S.",
    title = "{Ft values of the mirror {\ensuremath{\beta}} transitions and the weak-magnetism-induced current in allowed nuclear {\ensuremath{\beta}} decay}",
    eprint = "2109.08895",
    archivePrefix = "arXiv",
    primaryClass = "nucl-ex",
    doi = "10.1103/PhysRevC.107.015502",
    journal = "Phys. Rev. C",
    volume = "107",
    number = "1",
    pages = "015502",
    year = "2023"
}
\end{document}